\documentclass[MPS,Times1COL]{WileyNJDv5}

\usepackage{algorithm,algorithmicx}
\usepackage{bm}
\usepackage{framed}
\usepackage{comment}
\usepackage{subcaption}
\usepackage[utf8]{inputenc}
\DeclareUnicodeCharacter{FF0C}{,}

\articletype{ORIGINAL RESEARCH}%

\received{Date Month Year}
\revised{Date Month Year}
\accepted{Date Month Year}
\journal{Journal}
\volume{00}
\copyyear{2025}
\startpage{1}

\begin{document}

\title{Discrete-time Indirect Adaptive Control for Systems with Disturbances via Directional Forgetting: Concurrent Learning Approach}

\author[1]{Satoshi Tsuruhara}
\author[2]{Kazuhisa Ito}

\authormark{S. Tsuruahra \textsc{et al.}}
\titlemark{Discrete-time Indirect Adaptive Control for Systems with Disturbances via Directional Forgetting: Concurrent Learning Approach}

\address[1]{\orgdiv{the Graduate School of Engineering and Science, Functional Control Systems}, \orgname{Shibaura Institute of Technology}, \orgaddress{\state{307 Fukasaku, Minuma, Saitama 3378570}, \country{Japan}} and also is \orgdiv{a research fellow with Japan Society for the Promotion of Science}, \orgaddress{\state{Tokyo 1020083}, \country{Japan}}}

\address[2]{\orgdiv{the Department of Machinery and Control Systems}, \orgname{Shibaura Institute of Technology}, \orgaddress{\state{307 Fukasaku, Minuma, Saitama 3378570}, \country{Japan}}}

\corres{Corresponding author Satoshi Tsuruhara. \email{nb23110@shibaura-it.ac.jp}}

\presentaddress{307 Fukasaku, Minuma, Saitama 3378570, Japan}

\fundingInfo{This work was supported by JSPS Grant-in-Aid for JSPS Fellows Grant Number JP23KJ1923.}

\abstract[Abstract]{Recently, adaptive control systems with relaxed persistent excitation (PE) conditions have been proposed to guarantee true parameter convergence and improve the transient response.
However, in some cases, sufficient control performance and parameter convergence cannot be easily achieved, with stability demonstrated only under ideal conditions, such as the absence of disturbances and matching conditions required.
In this study, we propose a novel adaptive control method for discrete-time systems with disturbances, which is not under an ideal case, that combines directional forgetting and concurrent learning.
The proposed method does not require the PE condition, information on disturbances, unknown parameters, or matching conditions, and it guarantees uniformly ultimately bounded (UUB).
It was also theoretically demonstrated that the ultimate bound can be designed based on the forgetting factor, which is a design parameter.
In addition, the upper bound decreases with time step, which is independent of the system order and/or target trajectory due to forgetting factor.
This also implies stronger stability than a normal UUB.
Numerical simulation results illustrate the effectiveness of the proposed method.}


\jnlcitation{\cname{%
\author{Satoshi T.},
\author{Kazuhisa I.}}.
\ctitle{Discrete--time Indirect Adaptive Control for Systems with Disturbances via Directional Forgetting: Concurrent Learning Approach}
\cvol{2021;00(00):1--18}.}
\maketitle
\renewcommand\thefootnote{}

\renewcommand\thefootnote{\fnsymbol{footnote}}
\setcounter{footnote}{1}

\section{Introduction}\label{sec1}
Adaptive control has attracted a great deal of attention because it allows simultaneous updating of both model and controller parameters even when the system parameters are unknown, and it is known to work very effectively against structural uncertainty \cite{ioannou,tao}. 
It is also generally known that the adaptive system can guarantee the true convergence of the parameters and the exponential stability of the tracking error provided that the regressor satisfy the persistent excitation (PE) condition.
On the other hand, data-driven control, a method similar to adaptive control that does not require the model structure of the plant system and designs the controller from input/output (I/O) data only, has been widely proposed \cite{Survey_DDC1,Survey_DDC2}.
For example, direct data-driven tuning methods, such as fictitious reference iterative tuning/virtual reference feedback tuning\cite{FRIT,VRFT}, which are directly controller parameter tuning methods using fictitious/virtual signals, and data-based optimal control methods \cite{Survey_DDC3}, such as data enable predictive control based on Willems' fundamental lemma \cite{DeePC} or online reinforcement learning based on adaptive dynamic programming \cite{ADP}, have been widely studied. 
However, even in these methods, PE condition is essential for guaranteeing stability and designing highly versatile control systems.
Thus, the PE condition is important as an evaluation index for the quality of measurement data in terms of both adaptive control and data-driven control.
However, it is difficult in practical applications to satisfy this condition independently of the ideal target trajectory due to the constraints of the experimental rigs and control system drivers. Therefore, the relaxation of this condition is a crucial issue.

Recently, focusing on adaptive identification and its control systems, many adaptive control systems under relaxing the PE condition, which can guarantee stability and true value convergence, have been proposed \cite{survey_PE1, survey_PE2}.
The potential benefits of adaptive control, exemplified by approaches such as concurrent learning (CL) \cite{CL1,CL2,DCL}, composite model reference adaptive control \cite{CL-composite}, and dynamic regressor extension and mixing \cite{DREM} have been highlighted.
In particular, CL improves on the conventional normalized gradient (NG) algorithm in these methods by adding a correction term that accumulates important past I/O data, enabling easy convergence of parameters to true values without PE conditions.

In the presence of disturbances, the adaptive controller design is of great practical importance to prevent unexpected control performance degradation and instability, and this is no exception in CL.
In the NG algorithm for system identification \cite{tao,DCL2}, to guarantee the boundedness of the estimated parameters, the disturbance is required to have $L_2$ property, however, it would be a more strict condition than that of the boundedness of the disturbance.
Even if the $L_2$ property of the disturbance is satisfied, the error of the estimated parameters cannot be guaranteed uniformly ultimately bounded (UUB) unless the PE condition is satisfied.
Of course, the PE condition is necessary and sufficient for the NG algorithm to achieve true value convergence in the absence of disturbances.
On the other hands, the continuous-time CL-based adaptive controller is more robust against bounded noise than the $\sigma$-modification method, which is traditional robust adaptive control design, for aircraft applications and guarantees UUB \cite{CL3}; however, it is based on several strict assumptions such as the existence of positive invariant sets and matching condition.
In addition, UUB for unstructured uncertainty using a radial basis function neural network (RBFNN) or disturbance was proven in \cite{DCL2} for discrete-time CL adaptive identification, \cite{CL-AC} for continuous-time CL adaptive control, and in \cite{CL-composite2} for continuous-time composite model reference adaptive control.
However, \cite{DCL2} is only used for parameter identification, and \cite{CL-AC,CL-composite2} are not extended to discrete-time systems. In particular, \cite{CL-AC} requires information on the disturbance, and \cite{CL-composite2} requires a matching condition, which is effective for unstructured uncertainty using the RBFNN, but does not discuss the stability in the presence of output disturbances.
These extensions to discrete-time systems are of practically important; however, very little research has been conducted on them.

In this paper, inspired by the continuous-time CL method based on the forgetting factor for parameter jump systems without disturbances \cite{DF-CL}, we propose a novel discrete-time directional forgetting (DF)-based CL adaptive control system with bounded disturbances.
Note, however, that this study does not examine the validity of the method for time-varying systems such as parameter jumps but rather the introduction of a directional forgetting method, a type of forgetting factor, for disturbance attenuation.
The proposed control method guarantees, for the first time, UUB for discrete-time systems with disturbances that require neither matching conditions nor information on the disturbances.
In the absence of a disturbance, it achieves
exponential stability, unlike the conventional robust adaptive control. 
In practice, straightforward extensions of conventional discrete-time CL-based adaptive identification algorithms \cite{DCL2} and continuous-time composite adaptive control methods \cite{CL-composite} result in the ultimate bound that strongly depend on the system order or, in some cases, cannot even be defined.
Consequently, achieving sufficient control performance or rigorously guaranteeing stability is not always feasible.
On the other hand, it is theoretically shown that the proposed method is able to not only estimate the ultimate bound but also, as the time step ahead, the upper bound is more tightly and non-increasing with respect to the time step owing to the forgetting factor, which is the design parameter.
Therefore, high tracking control performance is expected, especially in steady-state response.
Numerical simulations compare the proposed method with two discrete-time CL adaptive control systems for unstable nonlinear affine systems.
The effects of disturbance, parameter errors, convergence of tracking errors, and the effect of the value of the forgetting factor are discussed, respectively.

The main contributions of this study are as follows: 1) a novel adaptive controller design for nonlinear affine systems is proposed based on a DF-based discrete-time CL.
To the author's knowledge, combining the DF and CL algorithms for dicrete-time system, the concept applied to disturbance suppression, and a proof of stability in the presence of disturbances is discussed for the first time in this paper.
Since most discussions focus on continuous-time systems, discussions on discrete systems are significant in practice.
2) The stability analysis discusses the UUB of the proposed method, which is expected to achieve higher tracking control performance in steady-state response than these of the two discrete-time CL methods because the upper bound becomes smaller as the time step is.
3) Numerical simulations show that the proposed method achieved higher control performance and robustness against disturbance than conventional methods. Then, applicable control classes are identified.
4) The relationship between the tracking error including convergence rate of parameters and the value of the forgetting factor is evaluated by numerical simulation.

The remainder of this paper is organized as follows.
First, Section 2 shows the preliminaries and problem formulation. 
Next, Section 3 describes the three types of CL algorithms and their extension to indirect adaptive control systems for discrete-time domain.
Section 4 discusses the stability analysis for Section 3, which includes the derivation of the ultimate bound and proof by the Lyapunov theorem.
The robustness of the system against disturbances is then evaluated by two numerical simulations in Section 5.
Finally, Section 6 presents conclusions and future works.

\section{Preliminalies and Problem Formulation}
\subsection{Preliminalies}
Let $\lambda_{\min}(\cdot)$ and $\lambda_{\max}(\cdot)$ denote the minimum and maximum eigenvalues, respectively.
Furthermore, we define $\bar{\lambda}=\{\lambda_{\max}(k)\in\mathbb{R}\ |\ \max_{k}\lambda_{\max}(k), ^\forall k\geq 0\}$ and $\bar{\eta}=\{\eta(k)\in\mathbb{R}\ |\ \max_{k}\eta(k),\ ^\forall k\geq 0\}$.
Moreover, let $\|\cdot\|$ and $\|\cdot\|_F$ denote Euclidean norm for vectors and Frobenius norm for matrices, respectively.

\begin{definition}\label{def:PE}
\cite{tao} The bounded vector signal $\phi(k)\in\mathbb{R}^n, n\geq 1$, is persistently exciting (PE) if there exist $\delta>0$ and $\alpha_{\phi}>0$ such that
\begin{align}
\sum_{k=\sigma}^{\sigma+\delta}\phi(k)\phi^T(k)\geq\alpha_{\phi}I,\ ^\forall \sigma\geq k_0.
\end{align}
\end{definition}

\begin{definition}\label{def:FE}
\cite{tao} The bounded vector signal $\phi(k)\in\mathbb{R}^n,\ n\geq 1$, is exciting over the time sequence set $\{\sigma_0,\sigma_0+1,\ldots,\sigma_0+\delta_\phi\},\ \delta_\phi>0,\sigma_0\geq k_0$, if for some $\alpha_\phi>0$ it holds that
\begin{align}
\sum_{k=\sigma_0}^{\sigma+\delta_\phi}\phi(k)\phi^T(k)\geq \alpha_{\phi}I
\end{align}
\end{definition}
\noindent Definition \ref{def:FE} is a weaker condition than the PE condition (Definition \ref{def:PE}) and it is called the finite-time excitation (FE) condition.

\subsection{Problem Formulation}
Consider the following nonlinear input affine SISO system.
\begin{equation}\label{ma:sys}
y(k+1)=f(k)+g(k)u(k)+w(k),
\end{equation}
where functions $f(k),\ g(k)$ denote unknown nonlinear functions with output $y(k)\in\mathbb{R}$, input $u(k)\in\mathbb{R}$, and disturbance $w(k)\in\mathbb{R}$. 
The following four assumptions were made:
\begin{assumption}
Functions $f(k)$ and $g(k)$ represent structured dynamics; that is, $f(k)\triangleq\theta_f^T\phi_f(k)\in\mathbb{R}$ and $g(k)\triangleq\theta_g^T\phi_g(k)\in\mathbb{R}$.
We define $\theta=[\theta_f^T,\ \theta_g^T]^T\in\mathbb{R}^n$ and $\phi(k)=[\phi_f^T(k),\ \phi_g(k)u(k)]^T\in\mathbb{R}^n$.
Then, (\ref{ma:sys}) can be expressed as follows:
\begin{equation}\label{ma:sys_ex}
y(k+1)=\theta^T\phi(k)+w(k).
\end{equation}
\end{assumption}
\begin{assumption}
The regressor vector $\phi(k)$ does not include $u(k)$ generated by the current time step $k$ and related signals.
However, the element of the power of the input, which is not the current time step $k$, can be considered, i.e. $u^2(k-1),\ u^3(k-1)y(k)$, etc.
\end{assumption}
\begin{assumption}\cite{DCL-AC}\label{ass:upper_g}
$g(k)$ is bounded away from zero for all $k\geq 0$.
Furthermore, there exists some known $\underline{g}>0$ such that $|g(k)|\geq\underline{g}$.
\end{assumption}

\begin{assumption}\label{ass:W}
The disturbance is bounded for all $k\geq 0$.
Thus, there exists a constant but unknown $W\geq 0$ such that
\begin{equation}
|w(k)|\leq W,\quad ^\forall k\geq 0.
\end{equation}
\end{assumption}

\begin{remark}
The proposed method needs only $\underline{g}$ for Assumption \ref{ass:upper_g}, and the upper constant but unknown value $W$ for Assumption \ref{ass:W} is only for the stability analysis.
Some robust adaptive controls based on dead-zone, switched $\sigma-$modification, and projection types \cite{tao,narendra2012stable} require information on upper bounds on disturbances or parameters, and improper design can lead to instability and performance degradation.
On the other hand, $\sigma-$modification algorithm does not require such parameter information, however, instead, it reduces less convergence rate of the parameters and does not guarantee zero error convergence even if in the absence of disturbances.
\end{remark}

\section{Proposed methods}
This section presents three discrete-time CL algorithms and discusses their differences in the data collection algorithms.
Next, the proposed method extended to indirect adaptive control systems is described.

\subsection{Concurrent learning estimation}
To identify the parameter vector $\theta$, we introduced the following model:
\begin{equation}
\hat{y}(k+1)=\hat{f}(k)+\hat{g}(k)u(k)=\hat{\theta}^T(k)\phi(k),
\end{equation}
where $\hat{\cdot}$ denotes the estimated value.
Let $\tilde{\theta}\triangleq\hat{\theta}-\theta\in\mathbb{R}^n$ be the parameter error.
Subsequently, we obtain the identification error $q(k+1)\in\mathbb{R}$ as follows:
\begin{align}\label{ma:re_qw}
q(k+1)&\triangleq \hat{y}(k+1)-y(k+1)\notag\\
&=\tilde{\theta}^T(k)\phi(k)-w(k)
\end{align}
Note that, when $w(k)\equiv0$, we can represent $q^\prime(k+1)=\tilde{\theta}^T(k)\phi(k)$ as the identification error.
Similarly, the identification error at time $k$ for the $i$th measured regressor vector is expressed as follows:
\begin{equation}
q_i(k+1) = \tilde{\theta}^T(k)\phi(i)-w_i(k)
\end{equation}
where $w_i(k)$ denotes disturbance for $\tilde{\theta}^T(k)\phi(i)$, which means a disturbance to the $i$-th column of recorded data.

The CL adaptive identification differs from conventional methods such as normalized gradient (NG) and normalized least means square (NLMS) in that it updates based on current time information $\phi(k)$ and recorded data.
The CL can guarantee parameter convergence with relaxed PE conditions, and global exponential stability of parameters is guaranteed under ideal conditions, such as no noise and/or disturbance.
The unique feature of adaptive identification is that it updates based on recorded data, not just information on the current time.
Several approaches have been proposed to collect and utilize this recorded data.

Firstly, we present the conventional CL method \cite{DCL}, which is known to be effective in the discrete-time domain as follows:
\begin{align}\label{ma:CL_tra}
\hat{\theta}(k+1) = \hat{\theta}(k) - \eta(k)\frac{\phi(k)q(k+1)}{m^2(k)}-\sum_{j=1}^{p}\eta(k)\frac{\phi(j)q_j(k+1)}{m^2(\phi(j))},
\end{align}
where $m(k)\triangleq \sqrt{\alpha+\phi^T(k)\phi(k)}$, $\alpha>0$.
$p$ denotes the number of columns of regressor vector to be recorded.
Let the recorded data matrix $Z\in\mathbb{R}^{n\times p}$, for some $p>n$, then $Z=[\phi_1/m(1),\ldots \phi_p/m(p)]$.
Moreover, $\eta(k)$ denotes the weight, and to guarantee stability for parameter error, we need to select following interval.
\begin{align}\label{ma:CL_weight}
  0<\eta<\frac{2m^2(k-1)}{2\lambda_{\max}(\phi(k)\phi^T(k))+\lambda_{\max}(ZZ^T)m^2(k-1)}
\end{align}
Moreover, as for the data collection algorithm, it has been shown that a high parameter convergence rate can be achieved by introducing an algorithm that sequentially swaps the current time data with the recorded data to maximize the inverse of condition number of the recorded data $ZZ^T\in\mathbb{R}^{n\times n}$.
The details are presented in \cite{DCL}.

On the other hand, as a counterpart, a pattern of collecting data in a continuous system is also considered \cite{CL-composite}.
In this paper, this discretized algorithm consists of the following parameter update law and algorithm.
\begin{align}\label{ma:CL}
\hat{\theta}(k+1) = \hat{\theta}(k) - \eta(k)\frac{\phi(k)q(k+1)}{m^2(k)}-\eta(k) (\Omega (k)\hat{\theta}(k)-M(k)), 
\end{align}
where $\Omega(k)\in\mathbb{R}^{n\times n}$ denotes the information matrix, and $M(k)\in\mathbb{R}^n$ denotes the auxiliary vector, which are updated using Algorithm \ref{alg:CL_basic}.
The algorithm updates both information matrix and auxiliary vector based on the regressor vector and measurement output when the rank condition is improved.
This is essential for ensuring that following Condition 1 is satisfied in the future time step.
Furthermore, if the last added regression vector $\phi_{\mathrm{last}}$ is far enough from the current regressor vector $\phi(k)$ via criteria with its threshold $\epsilon_{SM}$, both information matrix and auxiliary vector are updated regardless of the improvement in the rank condition, which is called stack-manager algorithm.
This is expected to improve the parameter convergence rate.
Note that the weights $\eta(k)$ can be similar by changing $ZZ^T$ in (\ref{ma:CL_weight}) to $\Omega(k)$.

\begin{algorithm}
\caption{CL with stack-manager algorithm}\label{alg:CL_basic}
\begin{algorithmic}
\State{\textbf{Step 1}: Set $\Omega(0)=0_{n\times n}$，$M(0)=0_{n}$, and $\varepsilon_{SM}>0$.
\State{\textbf{Step 2}: \If{$\mathrm{rank}
(\Omega (k))<\mathrm{rank}\left(\Omega (k)+\frac{\phi(k)\phi^T(k)}{m^2(k)}\right)$ $\mathrm{or}$ $\frac{\|\phi(k)-\phi_{\mathrm{last}}\|}{\|\phi(k)\|}\geq\varepsilon_{SM}$}
\State{$\Omega(k+1)=\Omega(k)+\frac{\phi(k)\phi^T(k)}{m^2(k)}$}
\State{$M(k+1)=M(k)+\frac{\phi(k)y(k+1)}{m^2(k)}$}
\Else
\State{$\Omega(k+1)=\Omega(k)$}
\State{$M(k+1)=M(k)$}
\EndIf
\State{\textbf{Step 3}: Set $k\leftarrow k+1$ and go to \textbf{Step 2}}}}
\end{algorithmic}
\end{algorithm}

\noindent{\bf \textrm{Condition 1.}} \textrm{The recorded data $ZZ^T\in\mathbb{R}^{n\times n}$ or information matrix $\Omega(k)\in\mathbb{R}^{n\times n}$ composed by the regressor vector $\phi(k)$ is the column full rank.}\\

\noindent Note that to determine whether the FE condition is satisfied, it is sufficient to evaluate whether \textbf{Condition 1} is satisfied.

To improve robustness against disturbances and/or noise, we propose a CL for discrete-time systems introducing the forgetting factor.
Several such algorithms have been studied and organized in \cite{FF_survey}.
The most common algorithm is exponential forgetting, which is very effective in guaranteeing global exponential stability of the parameters provided that the PE condition for the regressor vector is satisfied. However, when the PE condition is not satisfied, the BIBO stability of the information matrix cannot be guaranteed, which may lead to instability under noise and/or disturbance.
On the other hand, the DF algorithm, which performs forgetting only in the direction of excitation, has been proposed \cite{DF}.
It is very effective method that guarantees the BIBO stability of the information matrix regardless of the PE condition.
The proposed method applies this DF algorithm.
Let the forgetting factor $\mu\in(0,1]$.
The information matrix and auxiliary vector are updated to satisfy the following Condition 1, as in the following algorithm:
\begin{algorithm}
\caption{Directional forgetting-based CL algorithm}\label{alg:DF-CL}
\begin{algorithmic}
\State{\textbf{Step 1}: Set $\Omega(0)=0_{n\times n}$，$M(0)=0_{n}$
\State{\textbf{Step 2}: \If{$\mathrm{rank}
(\Omega (k))<\mathrm{rank}\left(\Omega (k)+\frac{\phi(k)\phi^T(k)}{m^2(k)}\right)$}
\State{$\Omega(k+1)=\Omega(k)+\frac{\phi(k)\phi^T(k)}{m^2(k)}$}
\State{$M(k+1)=M(k)+\frac{\phi(k)y(k+1)}{m^2(k)}$}
\Else
\State{$\Omega(k+1)=\Omega(k)-\mu\frac{\Omega(k)\phi(k)\phi^T(k)}{\phi^T(k)\Omega(k)\phi(k)}\Omega(k)+\frac{\phi(k)\phi^T(k)}{m^2(k)}$}
\State{$M(k+1)=M(k)-\mu\frac{\Omega(k)\phi(k)\phi^T(k)}{\phi^T(k)\Omega(k)\phi(k)}M(k)+\frac{\phi(k)y(k+1)}{m^2(k)}$}
\EndIf
\State{\textbf{Step 3}: Set $k\leftarrow k+1$ and go to \textbf{Step 2}}}}
\end{algorithmic}
\end{algorithm}

\subsection{Indirect adaptive controller}
This study extends three adaptive identification algorithms shown in Section 3.1 to indirect adaptive control systems, known as adaptive linearization methods \cite{DCL-AC,APBC}.
In this study, the reference model is as follows:
\begin{equation}\label{ma:refmodel}
y_m(k+1)=-a_m y_m(k)+b_m r(k)
\end{equation}
where $y_m(k)\in\mathbb{R}$ denotes the reference model output; $a_m,b_m\in\mathbb{R}$ denotes the system parameters, which are determined by the designer to be stable, and $r(k)\in\mathbb{R}$ denotes the reference trajectory.
Let $e(k)\triangleq y(k)-y_m(k)\in\mathbb{R}$ be tracking error, and we can express $e(k+1)$ as follows:
\begin{align}
e(k+1)&=y(k+1)-y_m(k+1)\notag\\
&=f(k)+g(k)u(k)+w(k)-y_m(k+1)
\end{align}
Now, supposing that $f(k)$, $g(k)$ and $w(k)$ are known, the ideal controller can be designed as follows:
\begin{equation}
u_{\mathrm{ideal}}(k)=\frac{\gamma_e e(k)-f(k)-w(k)+y_m(k+1)}{g(k)},
\end{equation}
where $\gamma_e$ denotes the design parameter chosen within $0<|\gamma_e|<1$.
If $u(k)=u_{\mathrm{ideal}}(k)$, then the error dynamics can be expressed as
\begin{align}
e(k+1)=\gamma_e e(k)
\end{align}
Thus, the ideal controller achieves linearization for the closed-loop system in this case, and the tracking error approaches exponentially zero.
In the practical problem, we replace $f(k),\ g(k)$ with each estimated parameter $\hat{f}(k),\hat{g}(k)$, respectively, and does not consider disturbance $w(k)$.
The adaptive controller is expressed as follows:
\begin{equation}\label{ma:controller}
u_{p}(k)=\frac{\gamma_e e(k)-\hat{f}(k)+y_m(k+1)}{\hat{g}(k)},
\end{equation}
where $\hat{f}(k)$ and $\hat{g}(k)$ are obtained by the adaptive identification algorithm used in Section 3.1.
Note that stability is not assumed in this discussion and will be discussed later.
To perform the stability analysis in Section 4, we derive the relationship between the tracking error $e(k+1)$ and the identification error $q(k+1)$ as follows:
\begin{align}\label{ma:re_eq}
e(k+1)&=y(k+1)|_{u(k)=u_p(k)}-y_m(k+1)+\hat{g}(k)(u_p(k)-u_p(k))\notag\\
&=f(k)+g(k)u_p(k)+w(k)-\hat{g}(k)u_p(k)+\gamma_e e(k)-\hat{f}(k)\notag\\
&=\gamma_e e(k)-q(k+1)|_{u(k)=u_p(k)}
\end{align}
where $|_{u(k)=u_p(k)}$ denotes the substitution of $u_p(k)$ for $u(k)$.

\begin{remark}
Referring to \cite{DF-CL}, if $\underline{g}$ is sufficiently small, then Assumption 3 can be satisfied by introducing an dead zone on the denominator of the controller such that
\begin{equation}
\hat{g}(k)=\left\{
\begin{split}
& \hat{g}(k)\ \quad \mathrm{if}\ |\hat{g}(k)|\geq \underline{g}\\
& \mathcal{G}(\hat{g}(k))\underline{g}\ \quad \mathrm{if}\ |\hat{g}(k)|< \underline{g}
\end{split}
\right.  
\end{equation}
where
\begin{equation}
\mathcal{G}(\hat{g}(k))=\left\{
\begin{split}
1,\quad &\mathrm{if}\ \hat{g}(k)>0\\
1,\quad &\mathrm{if}\ \hat{g}(k)=0\ \mathrm{and}\ \hat{g}(k-1)>0\\
-1,\quad &\mathrm{if}\ \hat{g}(k)=0\ \mathrm{and}\ \hat{g}(k-1)<0\\
-1,\quad &\mathrm{if}\ \hat{g}(k)<0
\end{split}
\right.
\end{equation}
Here, this necessitates that $\hat{\theta}_g(0)\neq 0$
be initialized such that $\hat{g}(0)\neq 0$ since $\hat{g}(-1)$ is not defined.
Also note that (\ref{ma:re_eq}) relationship is always satisfied even when the above algorithm is performed.
\end{remark}
\section{Stability analysis}
In this section, we mainly show a stability analysis of a proposed method, the DF-CL-based indirect adaptive controller: first, a) the effect of disturbance on the third term of CL is evaluated and its ultimately uniform bounded value are estimated, next, b) the boundedness of the estimated parameters is shown based on the Lyapunov stability theorem. Then, c) the boundedness of the signals in the system is then shown based on the mathematical induction and proof by contradiction. Finally, d) the UUB is proved for the tracking and parameter errors.

\subsection{Ultimately uniform bounded for CL}
We evaluate the influence of the third term in the parameter error equation of CL algorithm for the disturbance $w(k)$.
Let $k_e\in\mathbb{N}$ as the time step when the rank condition is satisfied for the first time, after which positive definiteness is always guaranteed.
Note that the following Lemma 1 is based on the features of DF \cite{DF,DF2}.
\begin{lemma}\cite{DF2}
We assume that the information matrix $\Omega(k)$ satisfies the rank condition (see Condition 1).
For the DF algorithm, the following matrix $U(k)$ is bounded from above to below for some $\mu>0$, $^\forall k\in[k_e,\infty)$:
\begin{equation}
\mu I\leq \mu\frac{\Omega(k-1)\phi(k-1)\phi^T(k-1)}{\phi^T(k-1)\Omega(k-1)\phi(k-1)}\triangleq U(k)\leq I.
\end{equation}
\end{lemma}
\begin{proof}
See \cite{DF2}.
\end{proof}

\begin{theorem}
Consider the case where the adaptive parameter update law is (\ref{ma:CL}) and Algorithm \ref{alg:DF-CL}.
The dynamics for parameter error including disturbance corresponding to the third term of the update law for CL, $W_{\Omega\tilde{\theta}}(k)$ can be expressed via (\ref{ma:re_qw}) as follows: 
\begin{align}\label{ma:re_Omega}
W_{\Omega\tilde{\theta}}(k)&\triangleq\sum_{i=0}^{k-k_e-1}\left(\prod_{j=1}^{i}\tilde{U}(k-j)\right)\frac{\phi(k-i-1)w(k-i-1)}{m^2(k-i-1)}+ \prod_{j=1}^{k-k_e}\tilde{U}(k-j)\sum_{i=0}^{k_e-1}\frac{\phi(i)w(i)}{m^2(i)}
\end{align}
where $\tilde{U}(k)\triangleq I-U(k)$.
An upper bound exists on the constant of the term containing the disturbance in (\ref{ma:re_Omega}): $^\forall k\geq 0$, which can be expressed as follows:
\begin{align}
\|W_{\Omega\tilde{\theta}}(k)\|<\frac{1+\mu k_e(1-\mu)^{k-k_e}}{\mu\sqrt{\alpha}}W
\end{align}
The ultimate bound is at $k=k_e$ above and can be expressed as follows:
\begin{align}
\|W_{\Omega\tilde{\theta}}(k)\|<\frac{1+\mu k_e}{\mu\sqrt{\alpha}}W
\end{align}
\noindent If it is selected to satisfy $\mu>\frac{1}{k_e}$, the following result is obtained for the disturbance term as $k\to\infty$
\begin{equation}\label{ma:lim_W}
\begin{split}
\lim_{k\to\infty}\|W_{\Omega\tilde{\theta}}(k)\|<\frac{1}{\mu\sqrt{\alpha}}W
\end{split}
\end{equation}
\end{theorem}

\begin{proof}
The update law of the information matrix for CL immediately yields the following relationship using Algorithm \ref{alg:DF-CL}:
\begin{align}
\Omega(k)&=\tilde{U}(k-1)\Omega(k-1)+\frac{\phi(k-1)\phi^T(k-1)}{m^2(k-1)}\\
&=\tilde{U}(k-1)\tilde{U}(k-2)\Omega(k-2)+\tilde{U}(k-1)\frac{\phi(k-2)\phi^T(k-2)}{m^2(k-2)}+\frac{\phi(k-1)\phi^T(k-1)}{m^2(k-1)}\\
&=\prod_{j=1}^{k-k_e}\tilde{U}(k-j)\Omega(k_e)+\sum_{i=0}^{k-k_e-1}\left(\prod_{j=1}^{i}\tilde{U}(k-j)\right)\frac{\phi(k-i-1)\phi^T(k-i-1)}{m^2(k-i-1)}\\
&=\prod_{j=1}^{k-k_e}\tilde{U}(k-j)\Omega(0)+\sum_{i=0}^{k-k_e-1}\left(\prod_{j=1}^{i}\tilde{U}(k-j)\right)\frac{\phi(k-i-1)\phi^T(k-i-1)}{m^2(k-i-1)}+ \prod_{j=1}^{k-k_e}\tilde{U}(k-j)\sum_{i=0}^{k_e-1}\frac{\phi(i)\phi^T(i)}{m^2(i)}.
\end{align}
Similarly, $M(k)$ is also derived.
Noting that $\Omega(0)=0,\ M(0)=0$ and (2), the third term of CL can be expressed as
\begin{align}
\Omega(k)\hat{\theta}(k)-M(k)&=\sum_{i=0}^{k-k_e-1}\left(\prod_{j=1}^{i}\tilde{U}(k-j)\right)\frac{\phi(k-i-1)\left[\phi^T(k-i-1)\hat{\theta}(k)-y(k-i)\right]}{m^2(k-i-1)}\notag\\
&\qquad+\prod_{j=1}^{k-k_e}\tilde{U}(k-j)\sum_{i=0}^{k_e-1}\frac{\phi(i)\left[\phi^T(i)\hat{\theta}(k)-y(i)\right]}{m^2(i)}\notag\\
&=\sum_{i=0}^{k-k_e-1}\left(\prod_{j=1}^{i}\tilde{U}(k-j)\right)\frac{\phi(k-i-1){q}^{\prime}_{k-i-1}(k+1)}{m^2(k-i-1)}+\sum_{i=0}^{k-k_e-1}\left(\prod_{j=1}^{i}\tilde{U}(k-j)\right)\frac{\phi(k-i-1)w(k-i-1)}{m^2(k-i-1)}\\
&\quad+\prod_{j=1}^{k-k_e}\tilde{U}(k-j)\left(\sum_{i=0}^{k_e-1}\frac{\phi(i){q}^{\prime}_{i}(k+1)}{m^2(i)}+\sum_{i=0}^{k_e-1}\frac{\phi(i)w(i)}{m^2(i)}\right)\\
&=\Omega(k)\tilde{\theta}(k)+W_{\Omega\tilde{\theta}}(k)\label{ma:Oth}
\end{align}
We considered the disturbance generated by the parameter update law and evaluated it by taking its Euclidean norm as follows: 
$
\left\|\frac{\phi(k)}{m(k)}\right\|<1,\ \frac{1}{m(k)}\leq\frac{1}{\sqrt{\alpha}}.
$
In addition, we introduced the constant matrix $\tilde{U}_{\max}=I-U_{\lambda_{\min}}$ such that
$
U_{\lambda_{\min}}=\{U(k)\in\mathbb{R}^{n\times n}\ |\ \mathrm{arg}\min{\lambda_{\min}(U(k)}),\ ^\forall k \geq k_e\}.
$
Then,
\begin{align}\label{ma:norm_W_oth}
\|W_{\Omega\tilde{\theta}}(k)\|&=\left\|\sum_{i=0}^{k-k_e-1}\left(\prod_{j=1}^{i}\tilde{U}(k-j)\right)\frac{\phi(k-i-1)w(k-i-1)}{m^2(k-i-1)} + \prod_{j=1}^{k-k_e}\tilde{U}(k-j)\sum_{i=0}^{k_e-1}\frac{\phi(i)w(i)}{m^2(i)}\right\|\notag\\
&\leq\left\|\sum_{i=0}^{k-k_e-1}\left(\prod_{j=1}^{i}\tilde{U}(k-j)\right)\frac{\phi(k-i-1)w(k-i-1)}{m^2(k-i-1)}\right\|+\left\| \prod_{j=1}^{k-k_e}\tilde{U}(k-j)\sum_{i=0}^{k_e-1}\frac{\phi(i)w(i)}{m^2(i)}\right\|
\end{align}

We now consider the upper bound of $W_{\Omega\tilde{\theta}}(k)$ by dividing it into the following three cases; 1) $0\leq k_e\leq k_e-1$，2) $k_e\leq k$, 3) $k\to\infty$.

\begin{itemize}
  \item[1)] $0\leq k\leq k_e-1$\\
  In this interval, the column full rank of the information matrix is not satisfied because it is before time $k_e$. Note that forgetting is, therefore, not applied.
In this case, (\ref{ma:norm_W_oth}) can be rewritten as follows:
\begin{align}
\|W_{\Omega\tilde{\theta}}(k)\|&\leq\left\| \prod_{j=1}^{k-k_e}\tilde{U}(k-j)\sum_{i=0}^{k_e-1}\frac{\phi(i)w(i)}{m^2(i)}\right\|\\
&\leq\sum_{i=0}^{k_e-1}\left\|\frac{1}{m(i)}\right\|\left\|\frac{\phi(i)}{m(i)}\right\|\|w(i)\|< \frac{k_e}{\sqrt{\alpha}}W\label{ma:case_1}
\end{align}

  \item[2)] $k_e\leq k$\\
This interval is affected by forgetting factor due to time steps larger than time $k_e$. In this case, using $U_{\lambda_{\min}}=I-\tilde{U}_{\max}\geq\mu I$, which is a positive definite matrix, we can evaluate the upper bound as follows:
\begin{align}
\|W_{\Omega\tilde{\theta}}(k)\|&\leq\sum_{i=0}^{k-k_e-1}\|\tilde{U}^{i}_{\max}\|\left\|\frac{1}{m(k-i-1)}\right\|\left\|\frac{\phi(k-i-1)}{m(k-i-1)}\right\|\|w(k-i-1)\|\\
&\qquad + \left\|\prod_{j=1}^{k-k_e}\tilde{U}(k-j)\right\|\sum_{i=0}^{k_e-1}\left\|\frac{1}{m(i)}\right\|\left\|\frac{\phi(i)}{m(i)}\right\|\|w(i)\|\\
&<\left[\sum_{i=0}^{k-k_e-1}\|\tilde{U}^{i}_{\max}\|+k_e\|\tilde{U}_{\max}^{k-k_e}\|\right]\frac{W}{\sqrt{\alpha}}\\
&\leq\left[\left\|(I-\tilde{U}_{\max}^{k-k_e})U_{\lambda_{\min}}^{-1}\right\|+k_e\|(I-\mu I)^{k-k_e}\|\right]\frac{W}{\sqrt{\alpha}}\leq\frac{1+\mu k_e(1-\mu)^{k-k_e}}{\mu\sqrt{\alpha}}W\label{ma:case_2}
\end{align}
  \item[3)] $k\to\infty$\\
Taking the limit of (\ref{ma:case_2}), which can be expressed as follows:
\begin{align}
\lim_{k\to\infty}\frac{1+\mu k_e(1-\mu)^{k-k_e}}{\mu\sqrt{\alpha}}W=\frac{1}{\mu\sqrt{\alpha}}W
\end{align}
Compared to the upper bound on the interval of 1) $0\leq k\leq k_e-1$, we have
\begin{align}\label{ma:case_3}
\frac{1}{\mu\sqrt{\alpha}}W-\frac{k_e}{\sqrt{\alpha}}W=\frac{(1-\mu k_e)}{\mu\sqrt{\alpha}}W
\end{align}
For selecting the value of the forgetting factor so that $\mu>\frac{1}{k_e}$, we obtain the upper bound as $k\to\infty$,
\begin{align}\label{ma:kfin}
\lim_{k\to\infty}\|W_{\Omega\tilde{\theta}}(k)\|<\frac{1}{\mu\sqrt{\alpha}}W
\end{align}

\end{itemize}

\noindent Therefore, the constant upper bound exists for $\|W_{\tilde{\theta}\Omega}(k)\|$ $^\forall k\geq 0$.

Finally, the largest upper bounds in the three cases are evaluated to obtain the ultimate bound.
The largest upper bounds is considered to be obtained in the cases for $k=k_e-1$ and $k=k_e$, just before and after the forgetting installation.
Evaluating these differences for $k=k_e-1$ and $k=k_e$ using (\ref{ma:case_1}) and (\ref{ma:case_2}), we have
\begin{align}\label{ma:comp_ultimate}
\frac{1+\mu k_e}{\mu\sqrt{\alpha}}W-\frac{k_e}{\sqrt{\alpha}}W=\frac{1}{\mu\sqrt{\alpha}}W>0
\end{align}
\noindent Therefore, the ultimate bound is obtained as (\ref{ma:case_2}) at $k=k_e$.
\end{proof}

In the conventional method \cite{DCL2}, the ultimate bound is given as $\frac{p}{\sqrt{\alpha}}W,\ p\geq n$.
Thus, it depends on the number of recorded data considered by the designer.
Consider the case where the number of recorded data is chosen to be equivalent to the system order, i.e., $p=n$.
In this case, the ultimate bound is smaller for the conventional method, considering the result of (\ref{ma:comp_ultimate}).
However, the proposed method can reduce this upper bound as time $k$ increases due to the effect of the forgetting factor algorithm, yielding (\ref{ma:kfin}).
This implies that it does not depend on the system order or time step $k_e$, and the proposed method can estimate more tightly because of no dependency on the system order nor the number of recorded data.
It is also important to note that the upper bound expressed in (\ref{ma:case_2}) is a decreasing function with respect to time step $k$.
Conventional methods, on the other hand, as step time goes by, do not decrease this estimated value and always depends on the number of recorded data because without forgetting factor, its upper bound remains constant.
Therefore, the proposed method is expected to achieve higher control performance than conventional methods in the steady-state response.

Next, we compare the proposed method with the CL method using the stack-manager algorithm, which collects data according to the threshold $\epsilon_{SM}>0$ value, therefore, it is not possible to determine the ultimate bound as explained below. Because data is always accumulated, we cannot guarantee the UUB of the parameter error.
Even if the threshold $\epsilon_{SM}$ is sufficiently large, i.e., the data to be added are finite time, then, at least, this algorithm needs to collect the data so that the information matrix satisfies the column full rank to guarantee stability. In this case, the upper bound will be at least $\frac{k_e}{\sqrt{\alpha}}W$ and it is the same value as the ultimate bound of the proposed method. This implies that the CL method using the stack-manager algorithm does not makes the upper bounded smaller than the proposed method.

\begin{lemma}
Assume Condition 1 and weight $\eta(k)$ satisfying:
\begin{equation}
0<\eta(k)<\frac{2m^2(k)}{2\lambda_{\max}(\phi(k)\phi(k)^T)+\lambda_{\max}(\Omega(k))m^2(k)}
\end{equation}
The following Lyapunov function candidate is defined:
\begin{equation}\label{ma:V_theta}
V_{\tilde{\theta}}(k)=\tilde{\theta}^T(k)\tilde{\theta}(k)
\end{equation}
Subsequently, we obtain the following equation:
\begin{equation}\label{ma:derivation_Vtheta}
V_{\tilde{\theta}}(k+1)\leq \beta_1V_{\tilde{\theta}}(k)+2\tilde{\theta}^T(k)E(k)+e_w(k)-\beta_2\frac{{q}^{\prime 2}(k+1)}{m^2(k)}
\end{equation}
where
\begin{equation*}
\beta_1\triangleq 1-\eta(k)\lambda_{\min}(\Omega(k))\left(2-\frac{2\eta(k)\lambda_{\max}(\phi(k)\phi^T(k))}{m^2(k)}+\eta(k)\lambda_{\max}(\Omega(k))\right)
\end{equation*}
\begin{equation*}
\beta_2\triangleq \eta(k)\left(2-\frac{\eta(k)\lambda_{\max}(\phi(k)\phi^T(k))}{m^2(k)}\right)
\end{equation*}
\begin{equation*}
P(k)\triangleq I-\eta(k)\frac{\phi(k)\phi^T(k)}{m^2(k)}-\eta(k)\Omega(k)
\end{equation*}
\begin{equation*}
E(k)\triangleq \eta(k)P(k)\bar{\varepsilon}_w(k),\ e_w(k)\triangleq \eta^2(k)\bar{\varepsilon}_w^T(k)\bar{\varepsilon}_w(k),\ \bar{\varepsilon}_w(k)\triangleq \frac{\phi(k)w(k)}{m^2(k)}-W_{\Omega\tilde{\theta}}(k)
\end{equation*}
\noindent In addition, the UUB for parameter error is guaranteed.
\end{lemma}
\begin{proof}
From the update law for the CL (\ref{ma:CL}) and (\ref{ma:Oth}), we have
\begin{equation}
\begin{split}
\tilde{\theta}(k+1)&=\left[I-\eta(k)\frac{\phi(k)\phi^T(k)}{m^2(k)}-\eta(k)\Omega(k)\right]\tilde{\theta}(k)+\eta(k)\frac{\phi(k)w(k)}{m^2(k)} - \eta(k)W_{\Omega\tilde{\theta}}(k)\\
&=P(k)\tilde{\theta}(k)+\eta(k)\bar{\varepsilon}_w(k).
\end{split}
\end{equation}
Then, we have
\begin{align}
V(k+1) &= \tilde{\theta}^T(k+1)\tilde{\theta}(k+1)\\
&=[P(k)\tilde{\theta}(k)+\eta(k)\bar{\epsilon}_w(k)]^T\left[P(k)\tilde{\theta}(k)+\eta(k)\bar{\epsilon}_w(k)\right]\\
&=\tilde{\theta}^T(k)P^2(k)\tilde{\theta}(k)+\tilde{\theta}^T(k)P^T(v)\eta(k)\bar{\varepsilon}_w(k)+\eta(k)\bar{\varepsilon}^T_w(k)P(k)\tilde{\theta}(k)+\eta^2(k)\bar{\varepsilon}_w^T(k)\bar{\varepsilon}_w(k)\\
&=\tilde{\theta}^T(k)P^2(k)\tilde{\theta}(k) + 2\eta(k)\tilde{\theta}^T(k)P(k)\bar{\varepsilon}_w(k)+\eta^2(k)\bar{\varepsilon}_w^T(k)\bar{\varepsilon}_w(k)
\end{align}
Here, we focus on $\tilde{\theta}^T(k)P^2(k)\tilde{\theta}(k)$, we have
\begin{align}
\tilde{\theta}^T(k)P^2(k)\tilde{\theta}(k) &=\tilde{\theta}^T(k)\left[I-\eta(k)\frac{\phi(k)\phi^T(k)}{m^2(k)}-\eta(k)\Omega(k)\right]^2\tilde{\theta}(k)\\
&=\tilde{\theta}^T(k)\left[I-2\eta(k)\frac{\phi(k)\phi^T(k)}{m^2(k)}-2\eta(k)\Omega(k)+\eta^2(k)\frac{(\phi(k)\phi^T(k))^2}{m^4(k)}\right.\\
&\left.\quad+2\eta^2(k)\Omega(k)\frac{\phi(k)\phi^T(k)}{m^2(k)}+\eta^2(k)\Omega^2(k)\right]\tilde{\theta}(k)\\
&=-\eta(k)\left(2I-\frac{\eta(k)\phi(k)\phi^T(k)}{m^2(k)}\right)\frac{{q}^{\prime 2}(k+1)}{m^2(k)}\\
&\quad+\tilde{\theta}^T(k)\left[I+2\eta^2(k)\Omega(k)\frac{\phi(k)\phi^T(k)}{m^2(k)}-2\eta(k)\Omega(k)+\eta^2(k)\Omega^2(k)\right]\tilde{\theta}(k)\\
&\leq -\beta_2\frac{{q}^{\prime 2}(k+1)}{m^2(k)}+\tilde{\theta}^T(k)\left[I+2\eta^2(k)\Omega(k)\frac{\phi(k)\phi^T(k)}{m^2(k)}-2\eta(k)\Omega(k)+\eta^2(k)\Omega^2(k)\right]\tilde{\theta}(k)
\end{align}
Here, we can select $\eta(k)$ as follows:
\begin{align}\label{ma:eta_NG}
  0<\eta(k)<\frac{2m^2(k)}{\lambda_{\max}(\phi(k)\phi^T(k))}\triangleq\eta_{\mathrm{NG}}(k)
\end{align}
Then, we have $\beta_2>0$.
Moreover, we have
\begin{align}
\tilde{\theta}^T(k)P^2(k)\tilde{\theta}(k)&\leq \left[1-\eta(k)\lambda_{\min}(\Omega(k))\left(2-\eta(k)\left(\frac{2\lambda_{\max}(\phi(k)\phi^T(k))}{m^2(k)}+\lambda_{\max}(\Omega(k))\right)\right)\right]V(k)-\beta_2\frac{{q}^{\prime 2}(k+1)}{m^2(k)}\\
&=\beta_1V(k)-\beta_2\frac{{q}^{\prime 2}(k+1)}{m^2(k)}\\
\end{align}
Here, we can also select $\eta(k)$ as follows:
\begin{align}\label{ma:eta_CL}
0<\eta<\frac{2m^2(k)}{2\lambda_{\max}(\phi(k)\phi(k)^T)+\lambda_{\max}(\Omega(k))m^2(k)}\triangleq\eta_{\mathrm{CL}}(k)
\end{align}
Under Condition 1, we now confirm that $0<\beta_1<1$ for $\eta(k)$ above, and this is equivalent to show that $0<1-\beta_1<1$. To this end, we check if the following inequality.
\begin{equation*}
0<\eta(k)\lambda_{\min}(\Omega(k))\left(2-\frac{2\eta(k)\lambda_{\max}(\phi(k)\phi^T(k))}{m^2(k)}+\eta(k)\lambda_{\max}(\Omega(k))\right)<1
\end{equation*}
The lower side is trivial from the fact that $\eta(k)\lambda_{\min}(\Omega(k))> 0$.
On the other hand, noting $\Omega(k)>0$, let $a\triangleq\lambda_{\min}(\Omega(k))$, $b\triangleq\frac{2\lambda_{\max}(\phi(k)\phi^T(k))}{m^2(k)}+\lambda_{\max}(\Omega(k))$, consider the quadratic function $ab\eta^2(k)-2a\eta(k)+1$.
Let $D$ be the discriminant equation, we have
\begin{align}
\frac{D}{4}=a^2-ab=a(a-b),
\end{align}
here, $a-b$ can be expressed as follows:
\begin{align}
a-b&=\lambda_{\min}(\Omega(k))-\left(\frac{2\lambda_{\max}(\phi(k)\phi^T(k))}{m^2(k)}+\lambda_{\max}(\Omega(k))\right)\\
&=\frac{[\lambda_{\min}(\Omega(k))-\lambda_{\max}(\Omega(k))]m^2(k)-2\lambda_{\max}(\phi(k)\phi^T(k))}{m^2(k)} < 0
\end{align}
for $\phi(k)\neq \bm{0}_n$.
Since $D<0$, this function satisfies $ab\eta^2(k)-2a\eta(k)+1>0$.
Therefore, we have $0<\beta_1<1$ whenever $\eta(k)$ is selected in the range (\ref{ma:eta_CL}).
Moreover, we evaluate the magnitudes of (\ref{ma:eta_CL}) and (\ref{ma:eta_NG}).
\begin{align}
\bar{\eta}_{\mathrm{NG}}(k)-\bar{\eta}_{\mathrm{CL}}(k) &=\frac{2m^2(k)(\lambda_{\max}(\phi(k)\phi^T(k))+\lambda_{\max}(\Omega(k))m^2(k))}{\lambda_{\max}(\phi(k)\phi^T(k))(2\lambda_{\max}(\phi(k)\phi^T(k))+\lambda_{\max}(\Omega(k))m^2(k))}> 0
\end{align}
This implies that it is feasible when $\eta(k)$ is selected in the range (\ref{ma:eta_CL}).
Therefore, we can derive (\ref{ma:derivation_Vtheta}).

Furthermore, we evaluate the upper bound on the norm of $P(k)$, $E(k)$, $e_w(k)$, $\bar{\varepsilon}_w(k)$ to proof the boundedness of the parameter error.

\begin{align}
\|P(k)\|_F&\leq \|I_{n}\|_F +\bar{\eta}\left\|\frac{\phi(k)}{m(k)}\right\|^2+\bar{\eta}\|\Omega(k)\|_F\notag\\
&<\sqrt{n}+\bar{\eta}(1+\lambda_{\max}(\Omega(k))\sqrt{n}).
\end{align}
Note that, from (\ref{ma:eta_CL}), the value of the upper bound of $\eta(k)$ is bounded as follows.
\begin{align}
\frac{2}{\frac{2\lambda_{\max}(\phi(k)\phi^T(k))}{m^2(k)}+\lambda_{\max}(\Omega(k))}\in L_{\infty}
\end{align}
Thus, there exists $\bar{\eta}$.
From Theorem 1, we have
\begin{align}
\|\bar{\varepsilon}_w(k)\| &< \frac{W}{\sqrt{\alpha}}+W_{\Omega\tilde{\theta}}(k)\left|_{k=k_e}\right.\\
&=\frac{W}{\sqrt{\alpha}}+\frac{1+\mu k_e}{\mu\sqrt{\alpha}}W = \frac{1+(1+k_e)\mu}{\mu\sqrt{\alpha}}W
\end{align}
Therefore,
\begin{align}\label{ma:B_E}
\|E(k)\|&\leq\eta(k)\|P(k)\|_F\|\bar{\varepsilon}_w(k)\|\notag\\
&< \bar{\eta}(\sqrt{n}+\bar{\eta}(1+\lambda_{\max}(\Omega(k))\sqrt{n}))\frac{1+(1+k_e)\mu}{\mu\sqrt{\alpha}}W\triangleq\bar{B}
\end{align}
Moreover,
\begin{equation}\label{ma:B_Ew}
e_w(k)\leq\bar{\eta}^2\left(\frac{1+(1+k_e)\mu}{\mu\sqrt{\alpha}}W\right)^2\triangleq \bar{C}
\end{equation}
Thus, the difference in the Lyapunov function with respect to the parameter error can be expressed as
\begin{align}
\Delta V_{\tilde{\theta}}(k+1)\leq \bar{A}\|\tilde{\theta}(k)\|^2+2\bar{B}\|\tilde{\theta}(k)\|+\bar{C}
\end{align}
where $\bar{A}\triangleq \beta_1-1$.
\end{proof}
\begin{remark}
To improve the parameter convergence rate, $\beta_1$ should be close to 0.
To minimize $\beta_1$, substituting $\beta_1(\eta)$ to  $\beta=\beta_{\mathrm{CL}}(k)/2$, we have
\begin{align}\label{ma:convergence_rate}
\beta_1=1-\frac{\lambda_{\min}(\Omega(k))m^2(k)}{2\eta(k)\lambda_{\max}(\phi(k)\phi^T(k))+\lambda_{\max}(\Omega(k))m^2(k)}
\end{align}
Note that $\Omega(k)$ can be replaced as $ZZ^T(k)$.
Only $\Omega(k)$ in (\ref{ma:convergence_rate}) can be modified by the designer.
Therefore, maximizing the inverse of the condition number of the information matrix implies an improvement in the parameters convergence rate.
\end{remark}

\begin{lemma}
We define the following Lyapunov function candidate for tracking error:
\begin{equation}\label{ma:V_e}
V_e(k)=e^2(k).
\end{equation}
Then, we have
\begin{align}
V_e(k+1)\leq\beta_4V_e(k)+(1+\varepsilon)\left(1+\frac{1}{\varepsilon}\right){q}^{\prime 2}(k+1)+\left(1+\frac{1}{\varepsilon}\right)^2w^2(k+1),
\end{align}
where $\beta_3\triangleq 1-\gamma_e^2(1+\varepsilon)$, $\beta_4\triangleq 1-\beta_3=\gamma_e^2(1+\varepsilon)$.
\end{lemma}
\begin{proof}
Considering the difference in Lyapunov function candidates using the relationship between the identification and tracking errors (\ref{ma:re_eq}), we have
\begin{align}
\Delta V_e(k+1)&=e^2(k+1)-e^2(k)\notag\\
&=(\gamma_e e(k)-q(k+1))^2-e^2(k).
\end{align}
Using the relationships $^\forall a,\ b\in\mathbb{R}$ and $\varepsilon>0$, $(a+b)^2\leq(1+\varepsilon)a^2+(1+1/\varepsilon)b^2$ (see (13.52) in \cite{Inq}) twice, we note that the relationship between the identification error and disturbance (\ref{ma:re_qw}) yields
\begin{align}
\Delta V_e(k+1)&\leq \gamma_e^2(1+\varepsilon)e^2(k)+\left(1+\frac{1}{\varepsilon}\right)q^2(k+1)-e^2(k)\notag\\
&=-\beta_3V_e(k)+\left(1+\frac{1}{\varepsilon}\right)[{q}^{\prime}(k+1)-w(k)]^2\notag\\
&\leq -\beta_3 V_e(k)+(1+\varepsilon)\left(1+\frac{1}{\varepsilon}\right){q}^{\prime 2}(k+1)+\left(1+\frac{1}{\varepsilon}\right)^2w^2(k+1),
\end{align}
where $\varepsilon$ satisfying $0<\varepsilon<\frac{1-\gamma_e^2}{\gamma_e^2}$.
\end{proof}
\subsection{The boundedness of the signals in the closed-loop system}
Using the previous results, the boundedness of the signal in the system is proved by both the mathematical induction and contradiction.
Note that the proof technique was specially introduced in \cite{DESO-MFAC}.

\begin{lemma}
Consider the adaptive law in (\ref{ma:CL}) and the control law in (\ref{ma:controller}).
In this case, the regressor vector are bounded for all $k\geq 0$.
\end{lemma}
\begin{proof}
Let $k_f$ be a sufficiently large finite time step.
First, we proof that $\phi(k)\in L^{\infty},\ ^\forall k\in [0,k_f]$ using the mathematical induction.
For $k=0$, $y(0)$, $u(0)$ are bounded because these are the initial values, and $\phi(0)$ is clearly bounded.
Therefore, it is valid for $k=0$.
Next, suppose that the regressor vector $\phi(k_f-1)$ is bounded.
Since $\theta\in L_{\infty}$, using (\ref{ma:sys_ex}), we have
\begin{align}
&\phi(k_f-1),\ \theta\in L^{\infty}\Rightarrow y(k_f)\in L^{\infty}\\
&\phi(k_f-1),\ y(k_f)\in L^{\infty}\Rightarrow \phi_f(k_f)\in L^{\infty}
\end{align}
From Lemma 2, for any $k\geq 0$, $\hat{\theta}(k)$ is bounded.
Thus, 
\begin{align}
\phi_f(k_f),\ \hat{\theta}(k_f)\in L^{\infty}\Rightarrow \hat{f}(k_f)\in L^{\infty}
\end{align}
From adaptive control law in (\ref{ma:controller}) and Assumption 3 via Remark 2, we have
\begin{align}
\hat{f}(k_f), y(k_f),\ y_m(k_f),\ y_m(k_f+1)\in L^{\infty},\ \hat{g}(k_f)\neq 0 \Rightarrow u(k_f)\in L^{\infty}
\end{align}
Therefore, we have
\begin{align}
\phi(k_f-1),\ y(k_f),\ u(k_f)\in L^{\infty}\Rightarrow \phi(k_f)\in L^{\infty}
\end{align}
Thus, mathematical induction guarantees boundedness of the regressor vector for sufficiently large time steps.

Referring to \cite{DESO-MFAC}, we show boundedness for any $k\geq 0$.
At time step $k=k_f+1$, assume that $\phi(k_f+1)\notin L^{\infty}$
As shown above, $^\forall k\in[0,k_f]$, $\phi(k)\in L^{\infty}$.
Therefore, as with mathematical induction, $\phi(k_f+1)$ is bounded. This contradicts the assumption.
Therefore, the regressor vector $\phi(k)$ is bounded for any $k\geq 0$.
\end{proof}

\subsection{Proof of UUB}
\begin{theorem}
If Condition 1 and the following equation are satisfied, the CL-based adaptive control guarantees the following results:
\begin{equation}
0<\eta<\frac{2m^2(k)}{2\lambda_{\max}(\phi(k)\phi(k)^T)+\lambda_{\max}(\Omega(k))m^2(k)}\triangleq\bar{\eta}_{\mathrm{CL}}.
\end{equation}
\begin{itemize}
\setlength{\leftskip}{5mm}
  \item[(a)] without disturbance\\
  $\tilde{\vartheta}(k) = [\tilde{\theta}^T(k),e(k)]^T\to\bm{0}$ exponentially as $k\to\infty$
  
  \item[(b)] with disturbance\\
  $\tilde{\vartheta}(k) = [\tilde{\theta}^T(k),e(k)]^T$ is globally UUB.
\end{itemize}
\end{theorem}

\begin{proof}
We consider the following Lyapunov function candidate:
\begin{equation}
V(k)=\beta_5 V_{\tilde{\theta}}(k)+V_e(k)\\
\end{equation}
where $\beta_5$ will be defined later.
From (\ref{ma:V_theta}) and (\ref{ma:V_e}), we have
\begin{equation}
\min\left(\beta_5, 1\right)\|\tilde{\vartheta}(k)\|^2\leq V(k)\leq\max\left(\beta_5, 1\right)\|\tilde{\vartheta}(k)\|^2.
\end{equation}
From Lemmas 2 and 3, we obtain:
\begin{align}\label{ma:V_temp}
V(k+1)&\leq\beta_4V_e(k)+\beta_5\beta_1V_{\tilde{\theta}}(k)+2\beta_5\tilde{\theta}^T(k)E(k)+\beta_5e_w(k)\notag\\
&\quad -\beta_5\beta_2\frac{{q}^{\prime 2}(k+1)}{m^2(k)}+(1+\varepsilon)\left(1+\frac{1}{\varepsilon}\right){q}^{\prime 2}(k+1)+\left(1+\frac{1}{\varepsilon}\right)^2w^2(k+1)\notag\\
&=\beta_4 V_e(k)+\beta_5\beta_1V_{\tilde{\theta}}(k)+\frac{p(k){q}^{\prime 2}(k+1)}{m^2(k)}+2\beta_5\tilde{\theta}^T(k)E(k)+\beta_5e_w(k)+ \left(1+\frac{1}{\varepsilon}\right)^2w^2(k+1),
\end{align}
where
\begin{equation}
p(k)\triangleq m^2(k)(1+\varepsilon)\left(1+\frac{1}{\varepsilon}\right)-\beta_5\beta_2
\end{equation}
To satisfy $p(k)\leq 0$, we set $\beta_5$.
\begin{equation}
\beta_5\geq\frac{m^2(k)(1+\varepsilon)\left(1+\frac{1}{\varepsilon}\right)}{\beta_2}>0
\end{equation}
Here, from Lemma 4, regressor vector $\phi(k)$ is bounded for all $k\geq 0$.
Thus, for any time $k\geq 0$, there exists a finite $m(k)$, and we can take the positive constant $\beta_5$.

From (\ref{ma:V_temp}), we have:
\begin{align}\label{ma:upper_Lyap}
V(k+1)\leq\beta_6 V(k)+2\beta_5\tilde{\theta}^T(k)E(k)+\beta_5e_w(k)+ \left(1+\frac{1}{\varepsilon}\right)^2w^2(k+1)
\end{align}
where $\beta_6\triangleq\max(\beta_4,\ \beta_1)\in(0,1)$.
From the above, the upper bound of the Lyapunov function can be expressed in only two kind of terms: to reduce its function and to be affected by the disturbance.

For the case without a disturbance, from the above equation, we have
\begin{equation}
V(k+1)\leq\beta_6 V(k).
\end{equation}
Hence, the CL-based adaptive control guarantees exponential stability.

Next, we consider the case with disturbance.
From (\ref{ma:upper_Lyap}), we define $x_\vartheta\triangleq\|\tilde{\vartheta}\|$, then
\begin{align}\label{ma:A}
\Delta V(k+1)\leq A x_\vartheta^2(k) + 2\beta_5\|E(k)\|x_\vartheta(k)+\beta_5e_w(k)+ \left(1+\frac{1}{\varepsilon}\right)^2\|w(k)\|^2.
\end{align}
where $A\triangleq \beta_6-1<0$.
Next, we prove the bounds on $\|E(k)\|$ and $e_w(k)$.
Here, the following upper bounds hold using the proof of Lemma 2.
\begin{equation}\label{ma:B}
B \triangleq 2\beta_5\bar{B} \neq 0.
\end{equation}
and
\begin{equation}\label{ma:C}
C\triangleq\beta_5 \bar{C} +\left(1+\frac{1}{\varepsilon}\right)^2W^2 > 0.
\end{equation}
Therefore, by using (\ref{ma:A}), (\ref{ma:B}), and (\ref{ma:C}), we have
\begin{equation}
\Delta V(k)\leq A x^2_{\vartheta}(k)+Bx_{\vartheta}(k)+C\triangleq Q_E(k).
\end{equation}
Because $x_{\tilde{\vartheta}}(k)\geq 0$, $A<0$ and $C\geq 0$, there exists only one positive solution $\Theta$ to $Q_E(k)=0$, and the constant upper bound on the norm of $\tilde{\vartheta}$ can be expressed as follows:
\begin{equation}
\|\tilde{\vartheta}(k)\|\leq\max(\|\tilde{\vartheta}(0)\|,\ \Theta),
\end{equation}
\begin{equation}
\Theta\triangleq \frac{-B-\sqrt{B^2-4AC}}{2A}.
\end{equation}
If $x_{\tilde{\vartheta}}(k)\leq\Theta$, $\Delta V(k)<0$; otherwise, $\tilde{\vartheta}(k)$ remains within the positive invariant set \cite{DCL2,APBC}.
Therefore, it was determined to be UUB without requiring PE condition.
\end{proof}

\section{Nuemerical simulations}
This paper evaluates three cases: 
\begin{itemize}
 \setlength{\leftskip}{0.5cm}
  \item[$\quad$(A)] a system without disturbance
  \item[(B)] a system with disturbance which change rectangular shape
  \item[(C)] (B) above and data forgetting is introduced which is a design parameter on control performance
\end{itemize}
In addition to the proposed method, we compare discrete-time adaptive control systems with stack managers and data swapping algorithms and NG-baed adaptive control.

Consider the following unstable SISO discrete-time system, which is an extension of example in \cite{DCL-AC}.
\begin{align}
y(k+1)=-2y(k)+0.5y(k-1)+\exp\left(-\frac{\left(y(k)-\frac{\pi}{2}\right)^2}{4}\right)+0.8u(k)-0.1u(k-1)+w(k)
\end{align}
where $w(k)$ denote the disturbance and is set as follows:
\begin{equation*}
w(k)=\left\{
\begin{split}
&0.3,\quad k<500\\
&-0.3,\quad k\geq 500
\end{split}
\right.
\end{equation*}
The regressor vector and parameter vector can be expressed as follows:
\begin{align*}
\phi(k)&=[\phi_f^T(k)\ \phi_g(k)]^T=\left[y(k),\ y(k-1),\ \exp\left(-\frac{\left(y(k)-\frac{\pi}{2}\right)^2}{4}\right),\ u(k-1),\ u(k)\right]^T\\
\theta &=[\theta_f^T\ \theta_g]^T= [-2,\ 0.5,\ 1,\ -0.1,\ 0.8]^T
\end{align*}
In common with all cases, the system parameters of the reference model were set to be stable with $a_m=0.5$ and $b_m=0.5$ in (\ref{ma:refmodel}), the controller design parameter $\gamma_e=0.5$, and design parameters $\alpha=1$ in $m(k)$, respectively.
In addition, the initial parameter was set as $\theta(0)=[0,\ 0,\ 0,\ 0,\ 1]^T$ for Assumption 3 to be satisfied initially.
The value of the forgetting factor was set to $\mu=0.7$ for cases (A) and (B), and $\mu=0.1,\ 0.5,\ 0.7,\ 0.9$ for case (C).
Furthermore, we set the threshold $\varepsilon_{SM}=0.1$ for updating the information matrix for the stack-manager algorithm.

\begin{figure}[t]
\centering
\begin{minipage}[b]{1\linewidth}
    \centering
    \includegraphics[keepaspectratio, scale=0.25]{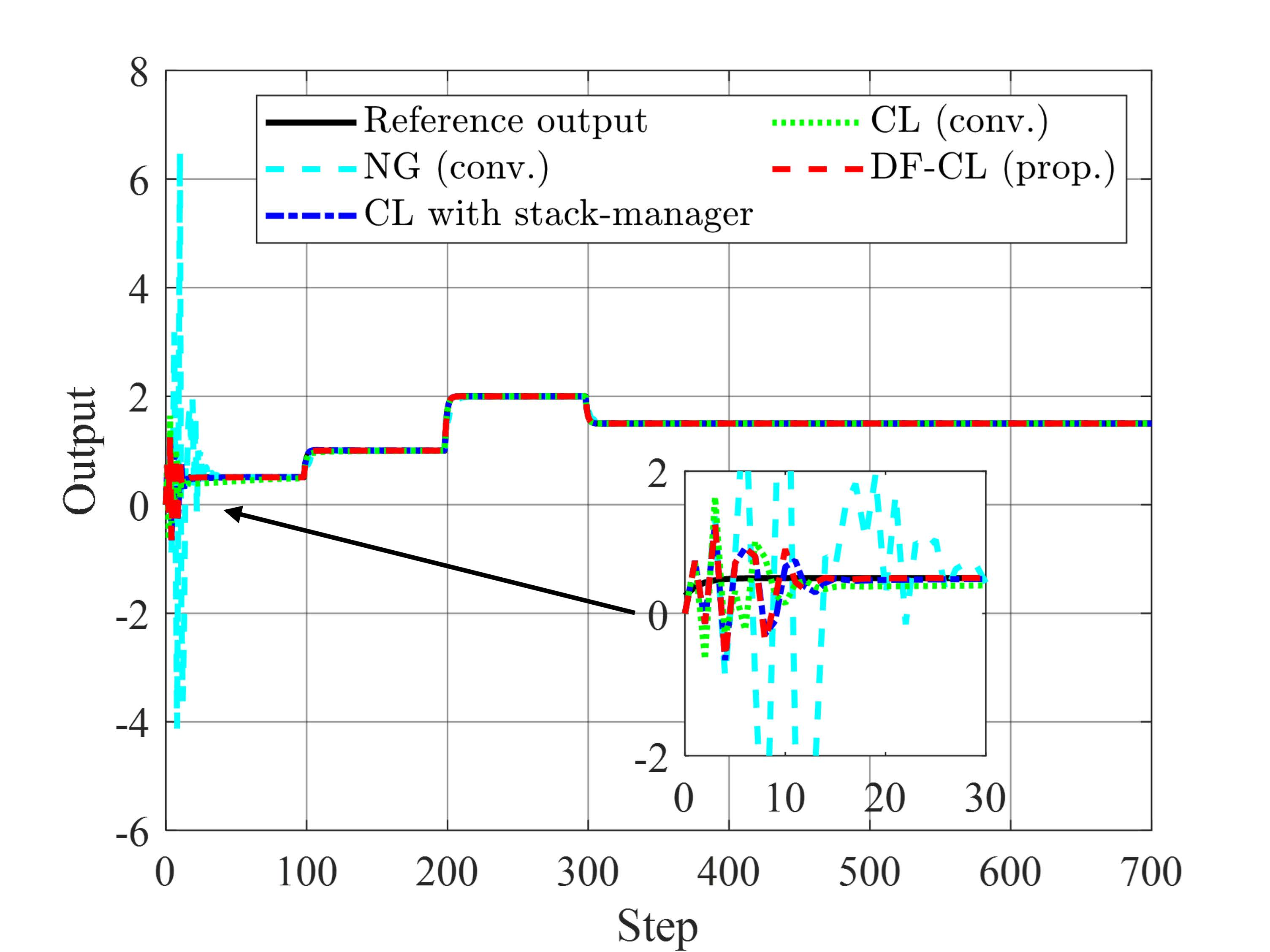} 
    \subcaption{Overall view}
    \centering
    \includegraphics[keepaspectratio, scale=0.25]{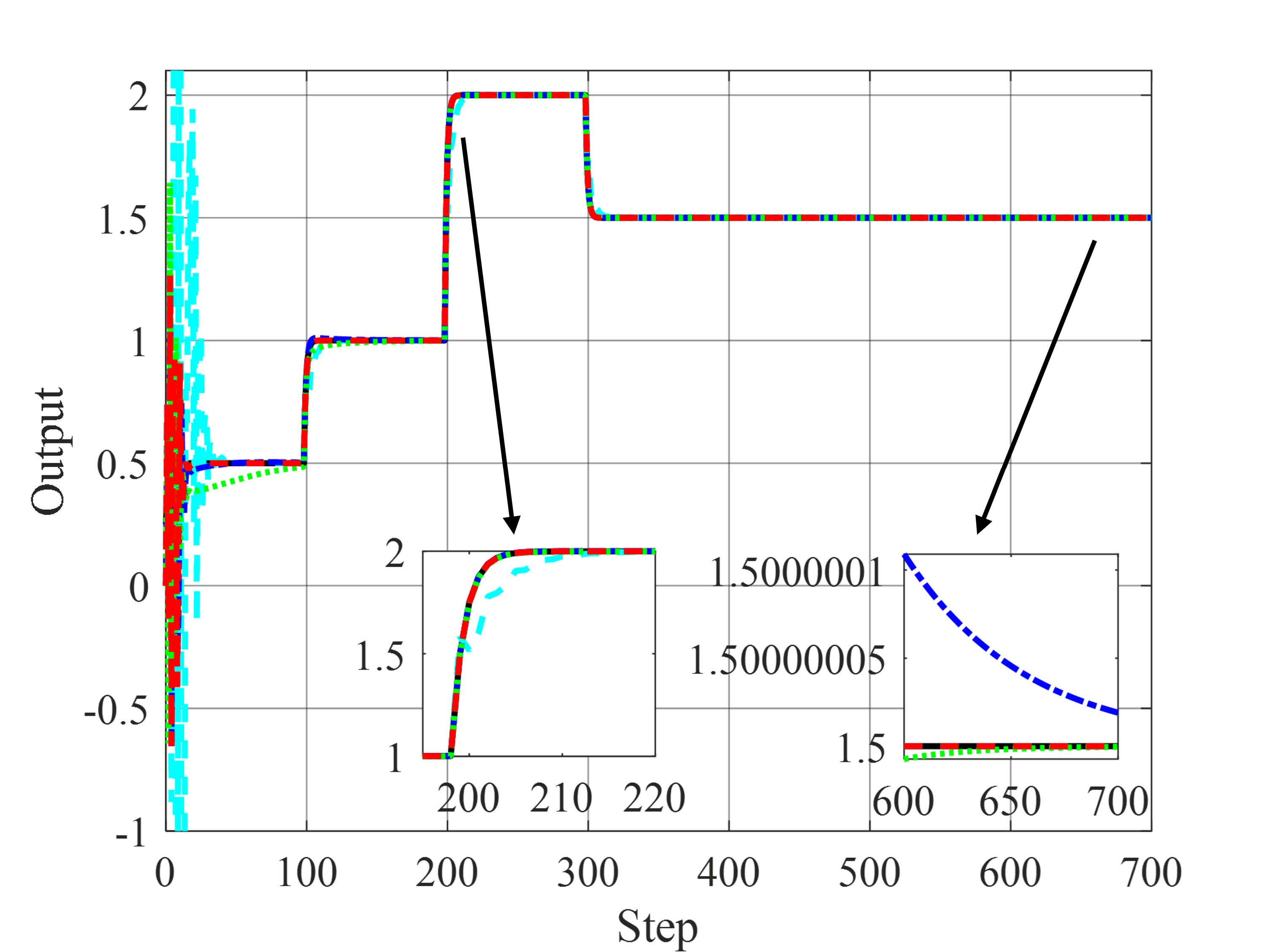}
    \subcaption{Enlarged view with y-axis restricted}
  \end{minipage}
  \caption{Comparison of the control performance in the absence of disturbance (Case (A))}
  \label{fig:comp_output_wo_dis}
\end{figure}

\begin{figure}[t]
\centering
\includegraphics[width=13cm]{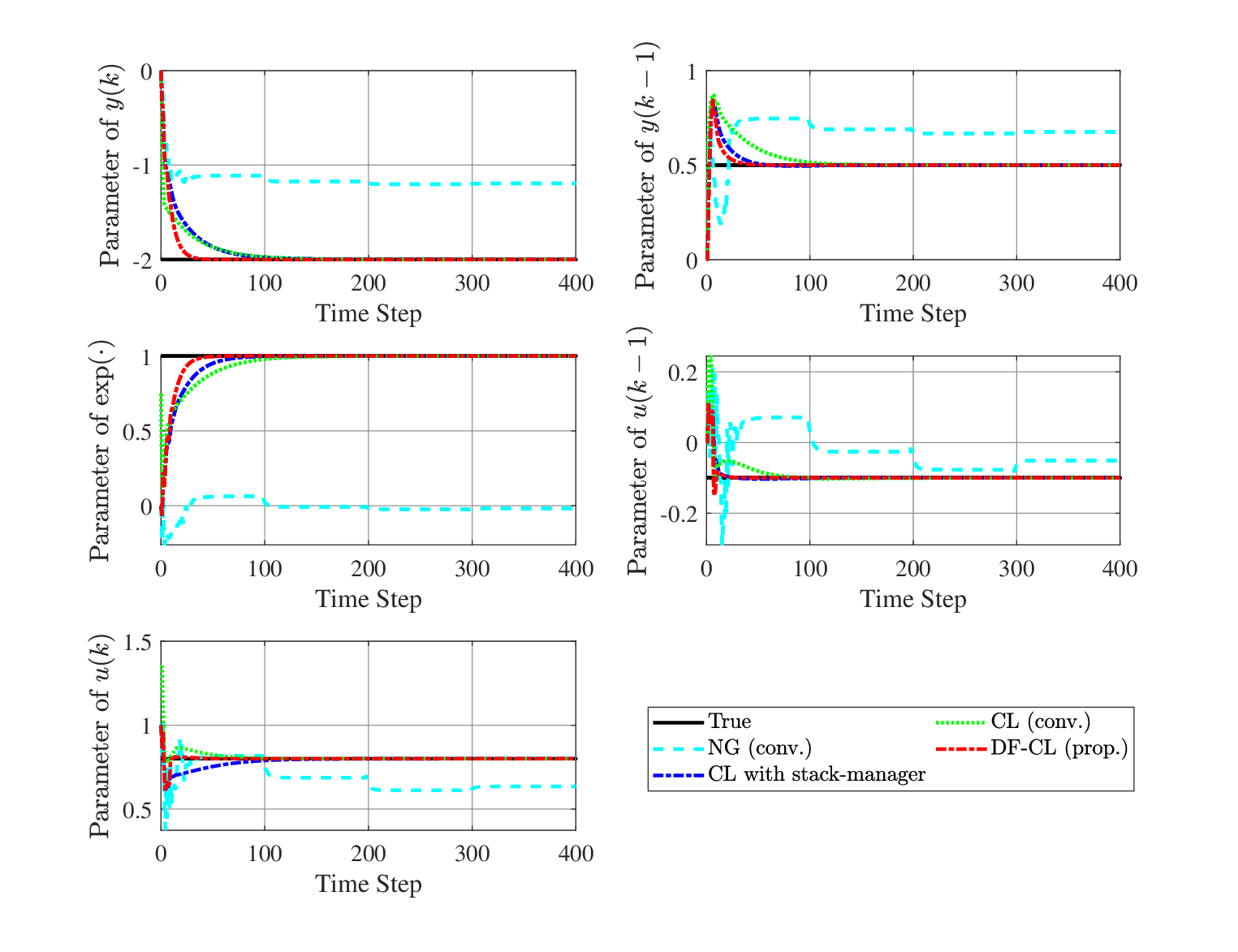}
  \caption{Comparison of the estimated parameters in the absence of disturbance (Case (A))}
  \label{fig:comp_parameter_wo_dis}
\end{figure}

\begin{figure}[t]
\centering
\includegraphics[width=13cm]{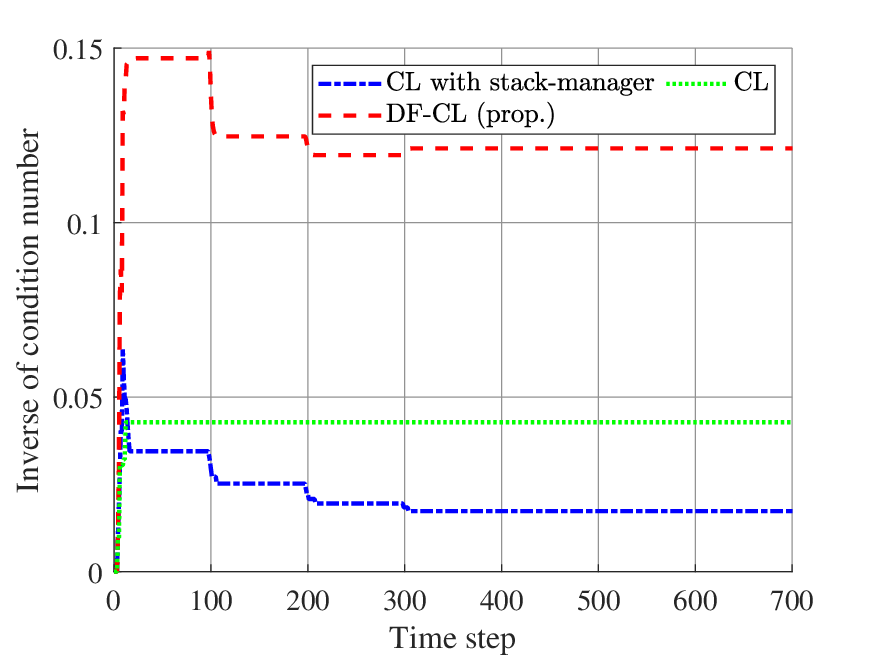}
  \caption{Comparison of the inverse of the condition number in the absence of disturbance (Case (A))}
  \label{fig:comp_cond_wo_dis}
\end{figure}

First, we evaluate (A) the case where no disturbance exists.
Figures \ref{fig:comp_output_wo_dis} and \ref{fig:comp_parameter_wo_dis} show the comparison of control performance and estimated parameters.
Figure \ref{fig:comp_output_wo_dis} shows that all methods achieve high control performance in steady-state response.
On the other hand, Fig. \ref{fig:comp_parameter_wo_dis} shows that the three CL-based methods achieve true value convergence, but the adaptive control system based on the NG method does not achieve true value convergence. This is because the PE condition is not satisfied, indicating that the CL method can achieve true value convergence without requiring the PE condition.
A comparison of the estimated parameters for the three types of CL methods shows that the proposed method achieves the fastest true value convergence of the parameters.
This can be attributed to the maximization of the inverse of the condition number shown in Fig. \ref{fig:comp_cond_wo_dis}.
The DF algorithm forgets the information matrix only in the direction of the excitation, thus maintaining a high inverse of the condition number by updating based on new data while minimizing the loss of past excitation information.
The CL with the stack-manager algorithm only adds data and updates the parameters with its data.
Thus, without forgetting data, the inverse of the condition number of the information matrix becomes smaller if the PE condition is not satisfied.
Moreover, although the CL with the data collection algorithm updates the information matrix to maximize the inverse of the condition number, the inverse of the condition number may not be sufficiently improved since it can only evaluate whether it has increased compared to the inverse of the condition number one step earlier.

The NG-based method initially exhibits a significant oscillatory response in the transient response of the control performance shown in Fig. \ref{fig:comp_output_wo_dis}. Furthermore, it cannot perfectly track the target trajectory during the transient interval in which it changes thereafter.
On the other hand, the proposed method achieves the fastest target trajectory tracking in the transient response when compared with the other two types of CL methods.
Combined with the control performance evaluation results based on the mean absolute error shown in Table 1, the proposed method achieves the highest control performance in transient and steady-state conditions, even without disturbance.

\begin{table}[t]
	\begin{center}
		\caption{Comparison results of control performance by mean absolute error}
		\label{table:case1}
		\begin{tabular}{ccccc}\hline
\rule[0mm]{0mm}{4mm} & NG (conv.) & CL with stack-manager & CL(conv.) & DF-CL (prop.)\\\hline\hline
\rule[0mm]{0mm}{4mm}w/o disturbance & 0.0589 & 0.0104 & 0.0183 & 0.00870\\
w/ disturbance & 0.0750 & 0.146 & 0.112 & 0.0169\\\hline
		\end{tabular}
	\end{center}
\end{table}

\begin{figure}[t]
\centering
\includegraphics[width=13cm]{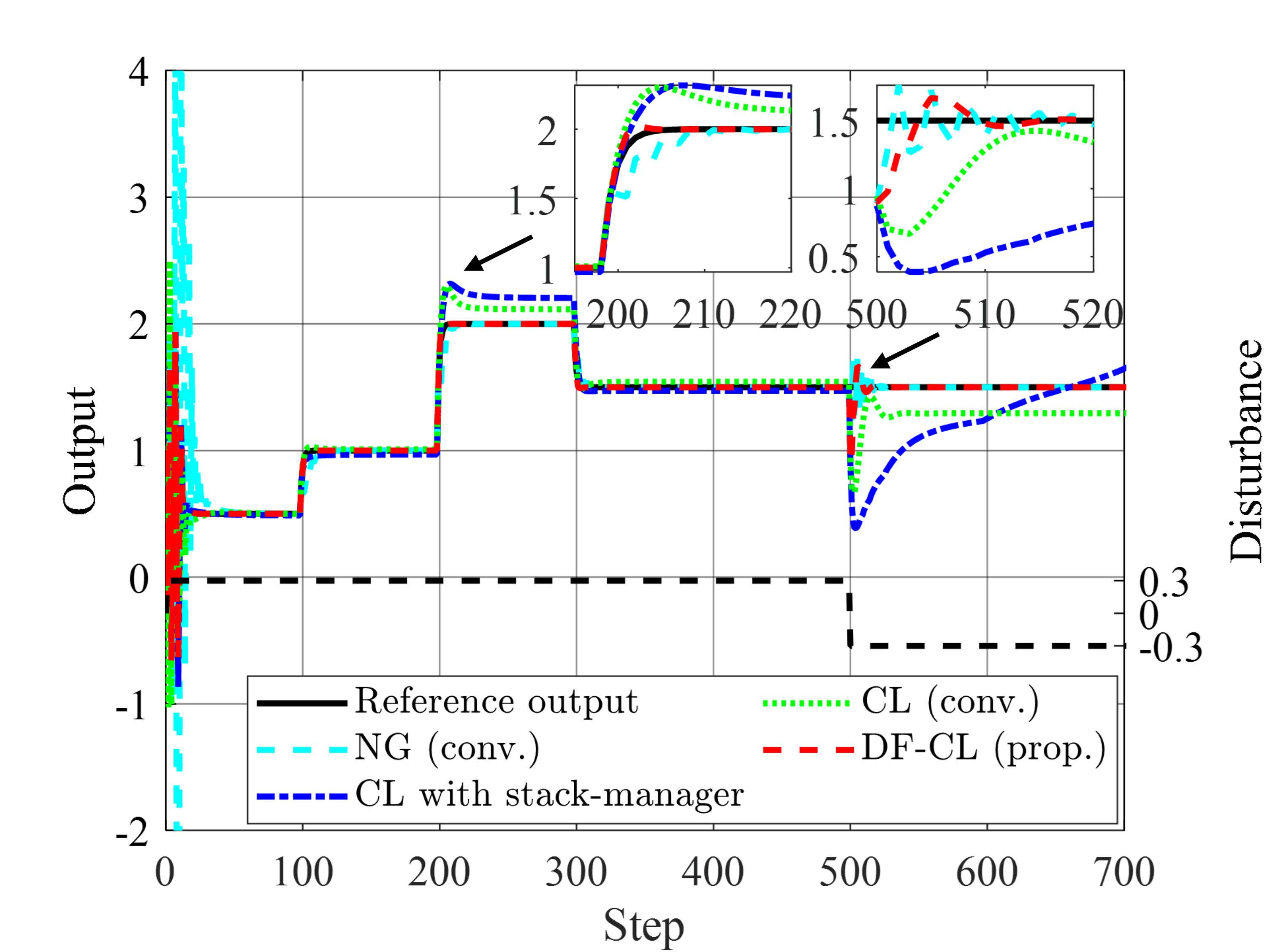}
  \caption{Comparison of control performance with disturbance (Case (B)), ($w(k)=0.3,\ k<500,\ -0.3,\ k\geq 500$)}
  \label{fig:comp_output_dis}
\end{figure}

\begin{figure}[t]
\centering
\includegraphics[width=13cm]{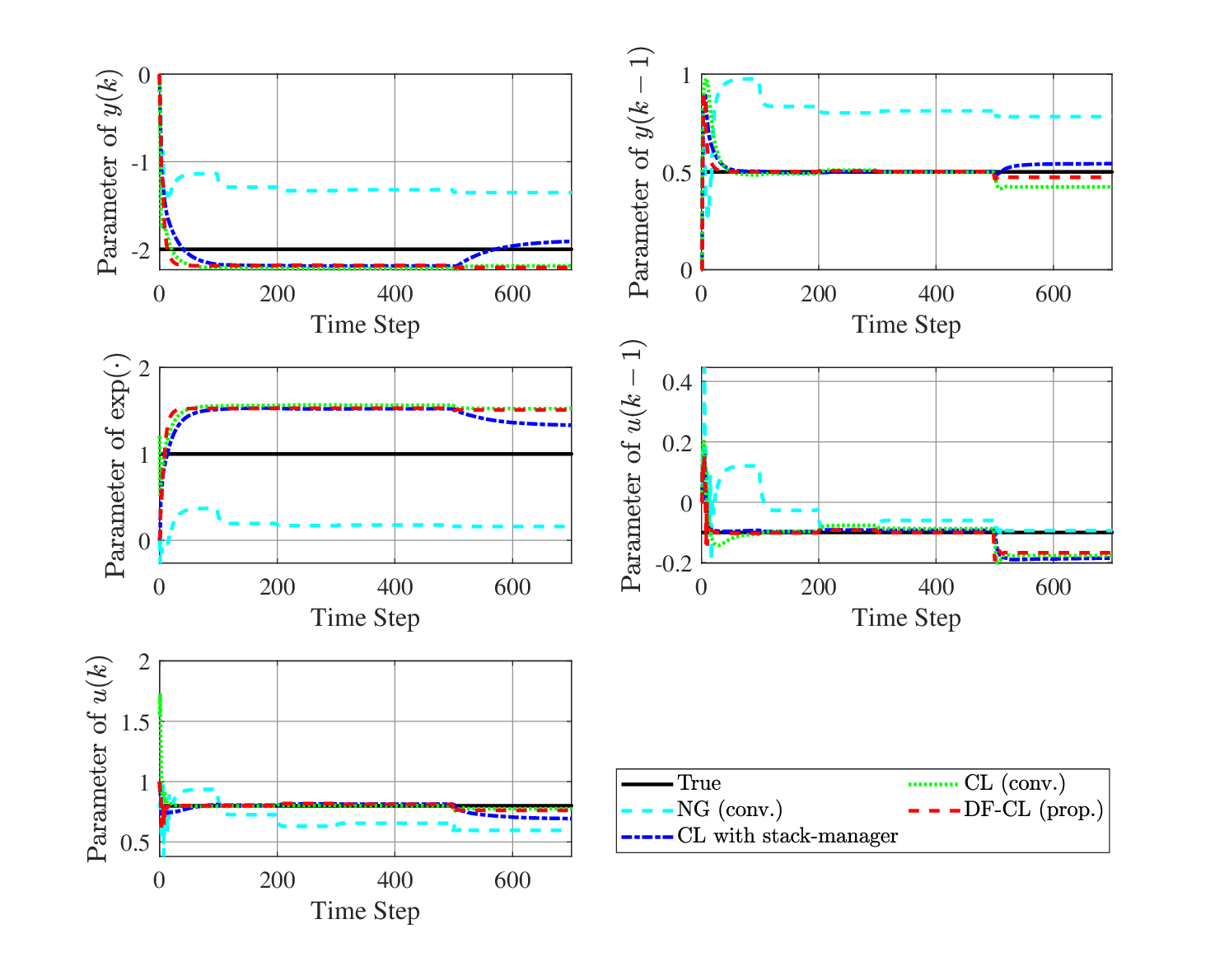}
  \caption{Comparison of estimated parameters with disturbance (Case (B)), ($w(k)=0.3,\ k<500,\ -0.3,\ k\geq 500$)}
  \label{fig:comp_parameter_dis}
\end{figure}

Next, we evaluate (B) the case where a disturbance exists and changes.
Figures \ref{fig:comp_output_dis} and \ref{fig:comp_parameter_dis} show a comparison of the control performance and parameter results.
Figure \ref{fig:comp_output_dis} shows that the proposed method has no steady-state deviation, no significant degradation in the transient response, and high robustness even when the value of the disturbance changes after 500 steps.
The NG-based algorithm degrades the transient response as in case (a) because the parameters shown in Figure 5 are far from the true values.
This method does not have recorded data and updates parameters for the data at the only current time.
Therefore, the effect of disturbance due to parameter errors is small and convergence to the true value is not forced to be achieved, resulting in high tracking control performance.

\begin{figure}[t]
\centering
\includegraphics[width=13cm]{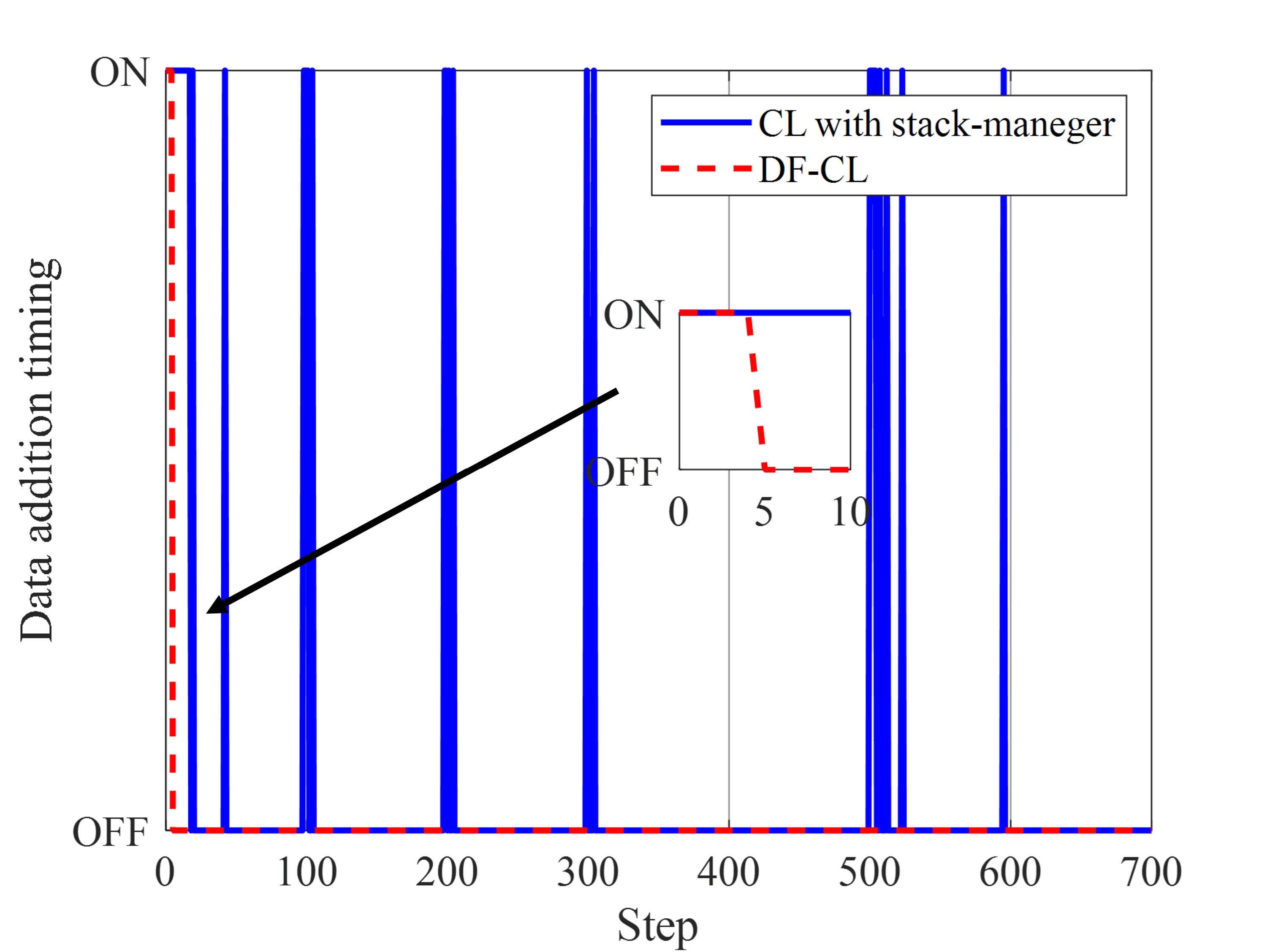}
  \caption{Comparison of data addition timing between CL with stack-manager and DF-CL (Case (B)), (Note that when data is added without forgetting, it is marked “On”.) }
  \label{fig:fig:comp_add_timing_dis}
\end{figure}

\begin{figure}[t]
\centering
\includegraphics[width=13cm]{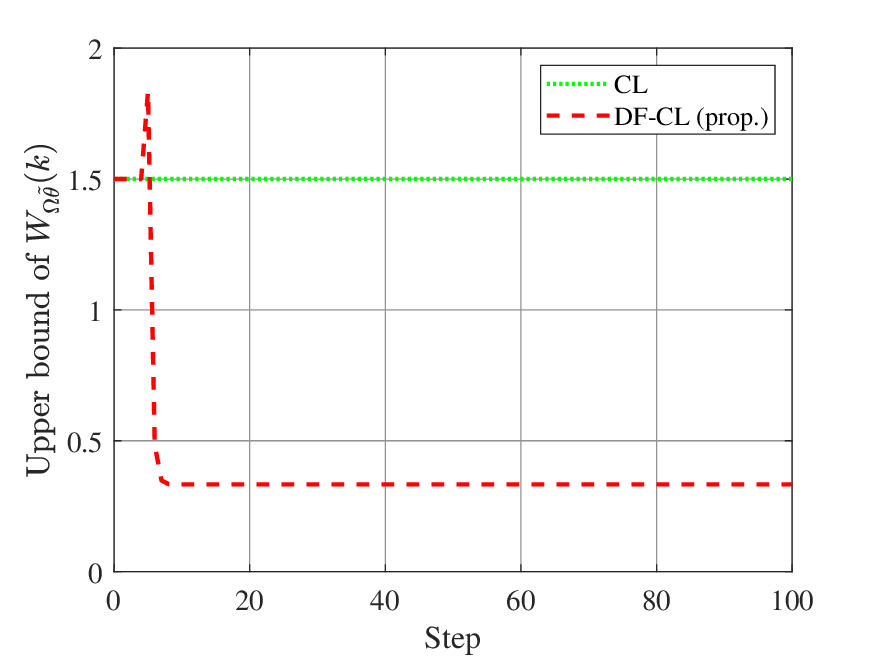}
  \caption{Time evolution of the upper bound of the disturbance influence in the third term of CL, which is in (\ref{ma:CL}), $\|W_{\Omega\tilde{\theta}}(k)\|$ via (\ref{ma:case_1}), (\ref{ma:case_2}) and (\ref{ma:case_3}) (comparison between CL(conv.) and DF-CL(prop.))}
  \label{fig:comp_time_upper}
\end{figure}

On the other hand, the two CL algorithms show that in the steady-state response, a steady-state deviation remains under the influence of the disturbance, indicating that they cannot suppress the influence of the disturbance, thus, they are low robustness against the disturbance.
The CL with stack-manager algorithm exhibits significant control performance degradation immediately after the disturbance changes, and it does not achieve sufficient tracking control.
This can be considered from the data addition timing shown in Fig. \ref{fig:fig:comp_add_timing_dis}.
This figure shows when the column full rank condition of the information matrix of CL is not satisfied or where the stack-manager algorithm adds data.
Note that data additions with forgetting are not shown.
The CL with the stack-manager algorithm adds, in particular, a huge number of data after around the $500$ step when the disturbance information changes. 
This method significantly degraded control performance after the disturbance change because the ultimate bound for the parameter error increases in dependence on the number of data additions.
On the other hand, since the proposed method only adds data in the first few steps without forgetting, the ultimate bound for the parameter error can be estimated much tighter than that of the conventional method.
For the conventional CL with data collection algorithm, as can be seen from Fig. \ref{fig:comp_time_upper}, the upper bound of the disturbance influence in the third term of CL, which is in (\ref{ma:CL}), $\|W_{\Omega\tilde{\theta}}(k)\|$ via (\ref{ma:case_1}), (\ref{ma:case_2}) and (\ref{ma:case_3}) remains at a fixed value regardless of the time step. 
Moreover, this value depends on the number of accumulated data.
On the other hand, the proposed method has a worse ultimate bound than this method. However, its upper bound is significantly reduced after the next step.
Although the decreasing speed depends on the value of the forgetting factor, it is a decreasing function concerning time $k$, and the proposed method is always smaller if it is selected to satisfy $\mu>\frac{1}{k_e}$ in Theorem 1.
Therefore, the proposed method achieves high tracking control performance and high robustness against disturbances.

\begin{figure}[t]
\centering
\includegraphics[width=13cm]{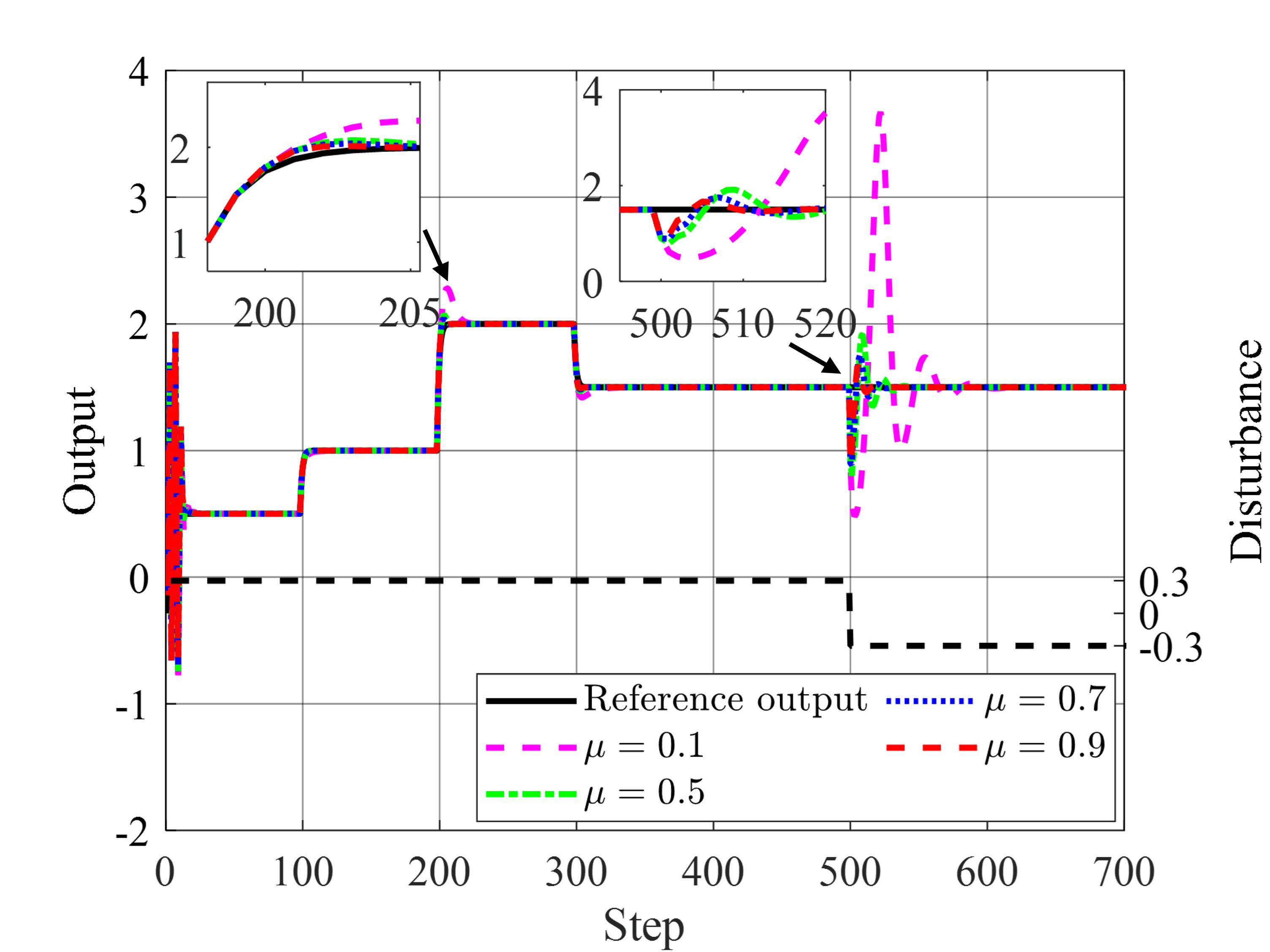}
  \caption{Comparison of control performance in DF-CL with changing the value of forgetting factors (Case (C)), ($\mu=0.1,\ 0.5,\ 0.7,\ 0.9$)}
  \label{fig:comp_output_FF}
\end{figure}

\begin{figure}[t]
\centering
\includegraphics[width=13cm]{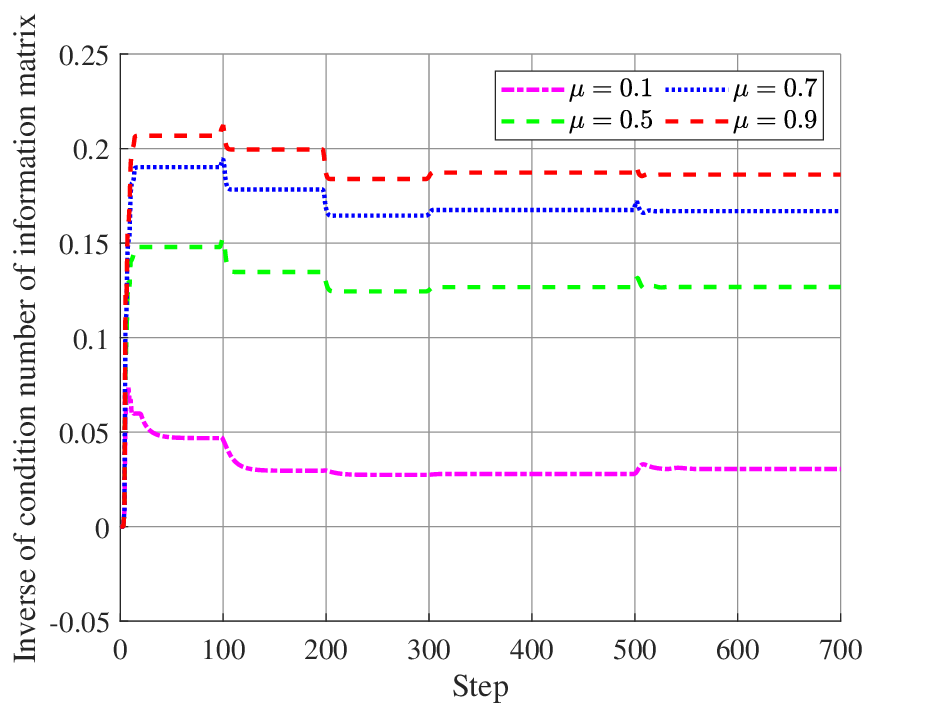}
  \caption{Comparison of condition number in DF-CL with changing the value of forgetting factors (Case (C)), ($\mu=0.1,\ 0.5,\ 0.7,\ 0.9$)}
  \label{fig:comp_cond_FF}
\end{figure}

Finally, we evaluate the case where (C) the effect of the forgetting factor.
Figure \ref{fig:comp_output_FF} shows the comparison of the control performance with changing the value of the forgetting factor.
The results show that control performance improves when the value of the forgetting factor is set closer to 1.
Note that unlike conventional forgetting factors, the closer this coefficient is to 1, the more data is forgotten.
This control performance result can be explained for two reasons.
The first is that the value of the forgetting factor closer to 1 provides the inverse of the condition number with increasing, as shown in Fig. \ref{fig:comp_cond_FF}.
This leads to a faster decrease in the Lyapunov function.
Note that changing the value of the forgetting factor closer to 1 does not necessarily improve the condition number but only tends to do so.
The second is that the ultimate bound for the parameter error becomes smaller as the value of the forgetting factor is close to 1.
This serves to suppress the growth of the Lyapunov function.
From the above, selecting the value of the forgetting factor close to 1 reduces the Lyapunov function, resulting in high control performance.

\section{Conculusion}
This paper proposed a novel discrete-time indirect adaptive control system for affine nonlinear systems, which combines the directional forgetting and the CL algorithm.
The proposed method can guarantee UUB for the parameter and tracking errors without the PE condition in the presence of disturbances.
Compared to the conventional method, the proposed method provides the upper bound decreases with time step due to forgetting factor.
This also implies stronger stability than a normal UUB.
Numerical simulations show that the proposed method achieves exponential true value convergence of the parameters and zero convergence of the error in the absence of disturbances. Its convergence rate is higher than the conventional method and the NG algorithm.
The proposed method also achieves high control performance and high robustness against disturbances in the presence of disturbances, both transient and steady-state.

\bibliography{mybibfile}

\begin{thebibliography}{10}
\providecommand{\url}[1]{{#1}}
\providecommand{\urlprefix}{URL }
\expandafter\ifx\csname urlstyle\endcsname\relax
  \providecommand{\doi}[1]{DOI~\discretionary{}{}{}#1}\else
  \providecommand{\doi}{DOI~\discretionary{}{}{}\begingroup \urlstyle{rm}\Url}\fi

\bibitem{ioannou}
Ioannou, P., Fidan, B.: Adaptive control tutorial.
\newblock SIAM (2006)

\bibitem{tao}
Tao, G.: Adaptive control design and analysis, vol.~37.
\newblock John Wiley \& Sons (2003)

\bibitem{Survey_DDC1}
Benosman, M.: Model‐based vs data‐driven adaptive control: An overview.
\newblock Int. J. Adapt Control Signal Process. 32(5), 753--776 (2018)

\bibitem{Survey_DDC2}
Hou, Z.S., Wang, Z.: From model-based control to data-driven control: Survey, classification and perspective.
\newblock Inf. Sci. 235, 3--35 (2013)

\bibitem{FRIT}
Kaneko, O.: Data-driven controller tuning: Frit approach.
\newblock IFAC Proceedings Volumes 46(11), 326--336 (2013).
\newblock \doi{https://doi.org/10.3182/20130703-3-FR-4038.00122}

\bibitem{VRFT}
Campi, M.C., Lecchini, A., Savaresi, S.M.: Virtual reference feedback tuning: A direct method for the design of feedback controllers.
\newblock Automatica 38(8), 1337--1346 (2002)

\bibitem{Survey_DDC3}
Tang, W., Daoutidis, P.: Data-driven control: Overview and perspectives.
\newblock In: 2022 American Control Conference (ACC), pp. 1048--1064, 2022

\bibitem{DeePC}
Coulson, J., Lygeros, J., D^^c3^^b6rfler, F.: Data-enabled predictive control: In the shallows of the deepc.
\newblock In: 2019 18th European Control Conference (ECC), pp. 307--312, 2019.
\newblock \doi{10.23919/ECC.2019.8795639}

\bibitem{ADP}
Farzanegan, B., Moghadam, R., Jagannathan, S., Natarajan, P.: Optimal adaptive tracking control of partially uncertain nonlinear discrete-time systems using lifelong hybrid learning.
\newblock IEEE Transactions on Neural Networks and Learning Systems 35(12), 17,254--17,265 (2024).
\newblock \doi{10.1109/TNNLS.2023.3301383}

\bibitem{survey_PE1}
Ortega, R., Nikiforov, V., Gerasimov, D.: On modified parameter estimators for identification and adaptive control. a unified framework and some new schemes.
\newblock Annual Reviews in Control 50, 278--293 (2020).
\newblock \doi{10.1016/j.arcontrol.2020.06.002}

\bibitem{survey_PE2}
Guo, K., Pan, Y.: Composite adaptation and learning for robot control: A survey.
\newblock Annual Reviews in Control 55, 279--290 (2023).
\newblock \doi{10.1016/j.arcontrol.2022.12.001}

\bibitem{CL1}
Chowdhary, G., Johnson, E.: Concurrent learning for convergence in adaptive control without persistency of excitation.
\newblock In: 49th IEEE Conference on Decision and Control (CDC), pp. 3674--3679, 2010

\bibitem{CL2}
Chowdhary, G., Yucelen, T., M^^c3^^bchlegg, M., Johnson, E.N.: Concurrent learning adaptive control of linear systems with exponentially convergent bounds.
\newblock Adaptive Control \& Signal 27(4), 280--301 (2013)

\bibitem{DCL}
Djaneye-Boundjou, O., Ord^^c3^^b3^^c3^^b1ez, R.: Parameter identification in structured discrete-time uncertainties without persistency of excitation.
\newblock In: 2015 European Control Conference (ECC), pp. 3149--3154, 2015.
\newblock \doi{10.1109/ECC.2015.7331018}

\bibitem{CL-composite}
Cho, N., Shin, H.S., Kim, Y., Tsourdos, A.: Composite model reference adaptive control with parameter convergence under finite excitation.
\newblock IEEE Transactions on Automatic Control 63(3), 811--818 (2018).
\newblock \doi{10.1109/TAC.2017.2737324}

\bibitem{DREM}
Ortega, R., Aranovskiy, S., Pyrkin, A.A., et~al.: New results on parameter estimation via dynamic regressor extension and mixing: Continuous and discrete-time cases.
\newblock on Automatic Control  (2020)

\bibitem{DCL2}
Djaneye-Boundjou, O., Ord^^c3^^b3^^c3^^b1ez, R.: Gradient-based discrete-time concurrent learning for standalone function approximation.
\newblock IEEE Transactions on Automatic Control 65(2), 749--756 (2020).
\newblock \doi{10.1109/TAC.2019.2920087}

\bibitem{CL3}
M^^c3^^bchlegg, M., Chowdhary, G., Johnson, E.: Concurrent Learning Adaptive Control of Linear Systems with Noisy Measurements.
\newblock \doi{10.2514/6.2012-4669}

\bibitem{CL-AC}
Zhao, Q., Duan, G.: Concurrent learning adaptive finite-time control for spacecraft with inertia parameter identification under external disturbance.
\newblock IEEE Trans. Aerosp. Electron. Syst. 57(6), 3691--3704 (2021)

\bibitem{CL-composite2}
Cho, N., Shin, H.S., Kim, Y., Tsourdos, A.: Composite model reference adaptive control under finite excitation with unstructured uncertainties.
\newblock In: 2023 62nd IEEE Conference on Decision and Control (CDC), pp. 529--535, 2023

\bibitem{DF-CL}
Lee, H.I., Shin, H.S., Tsourdos, A.: Concurrent learning adaptive control with directional forgetting.
\newblock IEEE Transactions on Automatic Control 64(12), 5164--5170 (2019).
\newblock \doi{10.1109/TAC.2019.2911863}

\bibitem{DCL-AC}
Djaneye-Boundjou, O., Ordonez, R.: Discrete-time indirect adaptive control of a class of single state systems using concurrent learning for parameter adaptation.
\newblock In: 2016 IEEE International Symposium on Intelligent Control (ISIC), pp. 1--6, 2016.
\newblock \doi{10.1109/ISIC.2016.7579988}

\bibitem{narendra2012stable}
Narendra, K.S., Annaswamy, A.M.: Stable adaptive systems.
\newblock Courier Corporation (2012)

\bibitem{FF_survey}
Lai, B., Bernstein, D.S.: Generalized forgetting recursive least squares: Stability and robustness guarantees.
\newblock IEEE Transactions on Automatic Control 69(11), 7646--7661 (2024).
\newblock \doi{10.1109/TAC.2024.3394351}

\bibitem{DF}
Cao, L., Schwartz, H.: A directional forgetting algorithm based on the decomposition of the information matrix.
\newblock Automatica 36(11), 1725--1731 (2000)

\bibitem{APBC}
Farrell, J.A., Polycarpou, M.M.: Adaptive approximation based control: Unifying neural, fuzzy and traditional adaptive approximation approaches.
\newblock John Wiley \& Sons (2006)

\bibitem{DF2}
Shin, H.S., Lee, H.I.: A new exponential forgetting algorithm for recursive least-squares parameter estimation.
\newblock arXiv preprint  (2020)

\bibitem{Inq}
Spooner, J.T., Maggiore, M., Ordonez, R., Passino, K.M.: Stable adaptive control and estimation for nonlinear systems: Neural and fuzzy approximator techniques.
\newblock John Wiley \& Sons (2004)

\bibitem{DESO-MFAC}
Chi, R., Hui, Y., Zhang, S., Huang, B., Hou, Z.: Discrete-time extended state observer-based model-free adaptive control via local dynamic linearization.
\newblock IEEE Transactions on Industrial Electronics 67(10), 8691--8701 (2020).
\newblock \doi{10.1109/TIE.2019.2947873}

\end{thebibliography}

\end{document}